# A Double Bind: Gendered Funding, Research Topics, and Academic Performance in the Social Sciences

Yang Ding [1, 2†], Ning Zhang [3†], Helen Bao [3†], Yu Jin [3†], Jiang Wu [4], Lianlian Wu [5, 6], Norman Weitemeier [7], Meng Huang [8], Alejandro Otazu Solorzano [9], Ana Paula Pineda Iriarte [10], Yunfeng Gao [11], Lok Man Michelle Tong [12, 13], Nancy Mukalayi [14], Pengfei Yin [15], Shuyu Hu [16], Yuxuan Xiao [17], Yarong Song [18], Jiajing Xu [1], Chenxu Li [19], Yi Bu [20,21,22,*]

[1] Business School, University of Edinburgh, Edinburgh EH8 9JS, United Kingdom

[2] Edinburgh Futures Institute, University of Edinburgh, Edinburgh EH3 9EF, United Kingdom

[3] Department of Land Economy, University of Cambridge, Cambridge CB3 9EP, United Kingdom

[4] School of Information Management, Wuhan University, Wuhan 430072, China

[5] Gastroenterology Department, Renmin Hospital of Wuhan University, Wuhan 430071, China

[6] Early Cancer Institution, University of Cambridge, Cambridge CB2 0XZ, United Kingdom

[7] Institute of International Business Law, Chair of Private Law, Philosophy of Law and Conflict of Laws, University of Münster, Münster 48161, Germany

[8] School for Business and Society, University of York, York YO10 5DD, United Kingdom

[9] School of Design, University of Edinburgh, Edinburgh EH3 9DF, United Kingdom

[10] School of Architecture Urban Planning Construction Engineering, Polytechnic University of Milan, Milano 20133, Italy

[11] School of Economics and Management, East China Normal University, Shanghai 200062, China

[12] Henley Business School, University of Reading, Reading RG6 6UD, United Kingdom

[13] Faculty of Finance, City University of Macau, Taipa, Macau Special Administration

Region of China

[14] Department of Economics, University of Pretoria, Pretoria 0002, South Africa

[15] School of Business, Trent University, Peterborough K9L 0G2, Canada

[16] Department of Ophthalmology, University of Cologne, Cologne 50937, Germany

[17] Amity Global Institute, Teesside University, Singapore 238853, Singapore

[18] Law School, Peking University, Beijing 100871, China

[19] Pratt School of Engineering, Duke University, Durham NC 27708, United States

[20] Department of Information Management, Peking University, Beijing 100871, China

[21] Center for Informationalization and Information Management, Peking University, Beijing 100871, China

[22] Center for Digital Intelligence Science and Education Research, Peking University Chongqing Research Institute of Big Data, Chongqing 401332, China

[†] Yang Ding, Ning Zhang, Helen Bao, and Yu Jin contributed equally to this paper.

* Corresponding Author: Yi Bu (buyi@pku.edu.cn)

**Acknowledgements:** Yang Ding acknowledges a Full PhD scholarship (#s2222886) from the University of Edinburgh; supported by the University of Edinburgh Development Trust Funds. Yi Bu acknowledges the National Natural Science Foundation in China (#72474009, L252400109).

# ABSTRACT

While female representation in social sciences is increasing, systemic gender disparities may persist in research funding and academic performance. Some argue that female scholars now receive equal opportunities, yet evidence suggests that gender imbalances remain, particularly in specific research areas. This study examines 12,945 National Science Foundation (NSF)-funded principal investigators in social sciences from 2000 to 2019 to assess gender disparities in grant allocation, research topics, and post-award academic performance. Findings reveal a dual imbalance. First, despite similar overall funding success rates, female scholars remain underrepresented in high-impact and traditionally male-dominated research topics. Male recipients are more represented in most funded topics, especially technology- and methodology-related ones, whereas female recipients are more concentrated in a smaller set of topics related to children, family, cognition, and health. Second, post-award performance patterns suggest that females outperform males in male-dominated fields, whereas males excel in female-dominated ones, undermining any presumed advantage of female scholars in their own research areas. These patterns may be associated with gendered constraints in academic career trajectories. Furthermore, early-career experiences shape these outcomes asymmetrically. In male-dominated topics, postdoctoral experience is associated with lower publication and citation performance for women but higher publication and citation performance for men. In female-dominated topics, postdoctoral experience is positively associated with women's publications and citations and with men's publication output. These findings suggest that policy discussions should consider not just overall funding equality, but also gendered disparities across research topics and career trajectories.

# INTRODUCTION

Around 80 years ago, 10,000 farm girls in Tennessee joined the Manhattan Project, performing the critical task of enriching uranium with calutrons—despite having only high school educations [1]. Remarkably, these "Calutron girls" outperformed male scientists and PhDs at UC-Berkeley [2]. Today, women are increasingly represented in research, particularly in the social sciences, where barriers are diminishing and female participation is rising [3]. However, equality in representation does not necessarily mean systemic biases are gone. For instance, while female Principal Investigators in U.S. social sciences have higher grant success rates, they receive, on average, over $10,000 less funding than males [4]. Understanding whether such disparities persist and how funding impacts female scholars' academic performance is crucial to supporting women in the field.

There is ongoing debate about whether gender barriers for female scholars in social sciences have been reduced. Some studies suggest that women still face conscious or unconscious discrimination, hindering their career progress, particularly in challenging or technical research areas [5–7]. Female researchers in fields like political science and economics also report less attention on their work compared to male scholars [4]. Furthermore, Antón et al. (2018) highlight the difficulties women face in high-impact research areas.

On the other hand, some argue that gender equality in social sciences has greatly improved, particularly regarding funding [9,10]. Government programs, e.g., ADVANCE in National Science Foundation (NSF), have increased funding opportunities for female scholars in both STEM and social sciences [11]. Female applicants for social science research grants are rising, with success rates approaching those of their male counterparts [12,13]. Furthermore, policies supporting early-career female researchers have led to a higher proportion of women in U.S. social sciences, surpassing male representation at all academic levels from 1998 to 2018 [14]. According to Ceci and Williams (2011), efforts to reduce discrimination have been largely successful, with funding differences attributed to selective behavior among scholars.

The polarized conclusions about gender inequality in social sciences stem from complex factors. Despite increased federal funding and specific legislation for female scholars, women's voices in academia remain weak, partly due to imbalances in funding allocation across research areas. For example, while the proportion of U.S. female doctors is rising, they are often concentrated in fields like Pediatrics and Obstetrics, while male-dominated fields remain less accessible [15]. In social sciences, female representation is strong in fields like Psychology, Economics, and Political Science, but quite poor in sub-fields like Cognitive Neuroscience [4]. Lundberg and Stearns (2019) find that female representation in Economics has stalled, with many women either choosing less male-dominated areas or leaving academia. As female inclusion grows in certain areas, it is often easier for women to enter specific fields than to gain equal access across all disciplines [17]. Thus, academic policymakers could consider not only overall funding equality but also topic-level representation across research areas [18].

The imbalance in support for female scholars in social science research may be obscured by data showing women are more represented in the field, such as higher overall female funding rates. This can create a misleading impression about the survival of women in academia [19].

While females and males in social sciences appear to have similar funding success rates, the data often overlooks specific research topics, where gender disparities may be more pronounced. Without detailed data on research topics, it would be unclear how well female scholars perform in particular areas, and overall funding rates may mask gender differences in sub-fields, potentially hindering career advancement [20]. There is limited evidence on whether women face gender inequalities in research topics, raising the need to examine if social science funding is gender-selective by research area to better understand whether topic-level concentration is associated with unequal academic outcomes.

Exploring the academic performance of female scholars after they have received funding may contribute to our further understanding of the current representation of female scholars in different gender-dominated research fields. Prior research has documented associations between research funding, promotion, and research impact [21]; uncovering the strengths and limitations of female scholars' academic performance across different research topics in social sciences could offer academia and policymakers more empirical evidences to better understand potential factors associated with female scholars' attrition from the science pipeline early in the process of enhancing their academic competitiveness [22]. This study examines 12,945 principal investigators (8,932 male and 4,013 female) funded by NSF in social sciences from 2000 to 2019 and assesses gender imbalances in research grant distribution across programs and topics. We identify which topics are dominated by males or females and investigate potential endogeneity in grant allocation. Using instrumental variables and the two-stage least-square (2SLS) model, we analyze differences in academic performance between male and female scholars post-funding. We also compare their performance in gender-dominant versus opposing research areas over time. Considering that the early career experiences of male and female PIs significantly shape their academic career trajectories, we examined how their early career experiences prior to applying for research funding influence the impact of the funding on their academic performance [23]. See **Supplementary Note 1** for additional insights into how gender and funding topics influence the science pipeline.

# MATERIALS AND METHODS

## Data

We examine scholars funded by NSF's SBE between 2000 and 2019 and their publication data. We focus on PIs from universities rather than firms or unaffiliated institutions, as greater variation in work environments and systems may lead to large disparities in funding motivations and grant purposes among scholars from differing organizations. Metadata includes the scholar's name, project title, grant start and end dates, affiliation, grant amount, grant number, country, and grant description. We employ an algorithm to make gender judgments about grantees' names and nationalities [1]. The gender of the PIs is considered in a binary way, without interfering with the scholars' own gender identifications and judgments. Ultimately, our data consisted of 12,945 PIs (8,932 males and 4,013 females), and 324,389 of their research articles indexed in the Microsoft Academic Graph (MAG) database and the Web of Science (WoS), two large-scale bibliographic databases. Trying to control for differences in publication efficiency and impact caused by research institutions on PIs as far as possible, we obtained publication records and citation counts for the research fields and universities to which the articles belonged through the research field labels and university information of the PIs' publications in the Elsevier literature database. Additionally, we match these universities to the Quacquarelli Symonds (QS) World University Rankings and the U.S. News & World Report list for university rankings and academic reputation scores. we confirmed the postdoctoral experience of 2,618 SBE-funded PIs through their ORCID profiles.

## Models

We construct a new treatment and control for all funded scholars. We use the academic performance of scholars who receive funding at a given stage to compare with scholars who have not yet received funding at that stage. We set the five years before and the five years after funding for scholars who received funding as the window between the untreated and treated stages. All scholars are funded. At this point, the control group refers to scholars who have not yet received funding in all years prior to the treatment group receiving funding and five years after funding. Using this comparison of treatment group and not-yet-receiving-treatment allows for some isolation of heterogeneity among the individuals being compared [24–26].

[1] https://gender-api.com/

We obtain the project categories of all PIs' funded projects in the SBE. SBE has released close to 1,000 research projects in the past, even though the names of these projects may be broadly similar. To obtain a clearer picture of the research topics of the research grants to which each PI belongs, we use the Latent Dirichlet Allocation (LDA) to perform funding topic extraction for all PIs' 12,945 research proposals. The LDA model and the specific extraction process for research funding topics are based on our prior research, which provides further details [27]. The classification of research funding topics is detailed in **Supplementary Note 4**.

Endogeneity issues are always faced when measuring research funding on scientific productivity, such as selection bias caused by the non-randomness of research funding decisions, reverse causality issues, etc. [28,29]. The results of the endogeneity test for SBE research grant allocations are provided in **Supplementary Note 5**. We treat funding as the focal explanatory variable and estimate its association with scholars' subsequent academic performance across gender and research topics. To address potential endogeneity in research funding, we use instrumental variables and the 2SLS model to measure the research funding effect,

$$F_{it} = \alpha_0 + \alpha_1 \boldsymbol{I}_{it} + \alpha_2 \boldsymbol{X}_{it} + \mu_{it} \quad (1)$$

$$P_{it} = \beta_0 + \beta_1 F_{it} + \beta_2 \boldsymbol{X}_{it} + \varepsilon_{it} \quad (2)$$

In the first-stage regression of the 2SLS model (Equation 1), we estimate the probability, $F_{it}$, of scholar $i$ receiving funding at time $t$. $\boldsymbol{I}$ is a vector of three instrumental variables, while $\boldsymbol{X}$ represents covariate variables. In the second stage (Equation 2), we estimate the impact of funding, the estimated value $\hat{F}_{it}$ from the first stage, on academic performance, $P$. The parameter of interest is $\beta_1$, indicating the coefficient of the funding effect. $\alpha_0$ and $\beta_0$ are constant terms; $\mu_{it}$ and $\varepsilon_{it}$ are residual terms. The instrumental variables used in this study follow our prior work and are intended to predict variation in the likelihood of receiving research funding [27]. Because the exclusion restriction cannot be directly tested, the 2SLS estimates should be interpreted as funding effects under the maintained IV assumptions. Additionally, to reduce concerns about reverse causality, we include the scholar's academic performance in the first three years of their academic career and a dummy variable marking that stage as the covariate in the equation [28]. The original first-stage regression results for all 2SLS models in this paper are presented in **Supplementary Note 6**. The instrumental variables used in this study are discussed in our prior research [27]. Additional diagnostics and discussion of the maintained IV assumptions are provided in **Supplementary Notes 7 and 8**.

# RESULTS

### Gender imbalance in research topics

Here in this paper, we examine scholars funded by NSF's Social Behavioral, and Economics Sciences (SBE) between 2000 and 2019 and their publication data. As shown in Fig. 1a, in the 30 categories with the greatest official number of grants announced by the SBE, 24 research programs have a proportion of male grantees that exceeds the overall officially certified male funding rate. Female scholars are only able to meet or exceed the overall female funding rate in six funded programs. Notably, in Neuroscience (Programs 1 and 4), Human Environment and Geographical Science (Programs 2 and 5), Economics, Decision making, Risk and Management Science (Program 3), Linguistics (Program 6), and Methodology, Measurements and Statistics (Program 7), the proportion of male recipients exceeded 80%, with only 0%-20% female recipients. In addition, funded programs closer to mathematics, engineering, data and information technology such as Secure and Trustworthy Cyberspace (Program 9), Mathematical Social and Behavioral Sciences (Program 12), Information Technology Research (Program 14), Data Infrastructure (Program 18), Science, Technology and Society (Program 21) and Biological Anthropology (Program 22), all have a lower percentage of females funded than the overall female funding rate. The programs where females are dominant tend to be associated with Facility Building (Programs 29 and 30) and Perception, Action and Cognition (Program 28). This further supports the findings of Casad et al. (2022) and Lundberg and Stearns (2019), who state that females in social sciences are under-represented in certain sub-research areas such as Neuroscience and Economics.

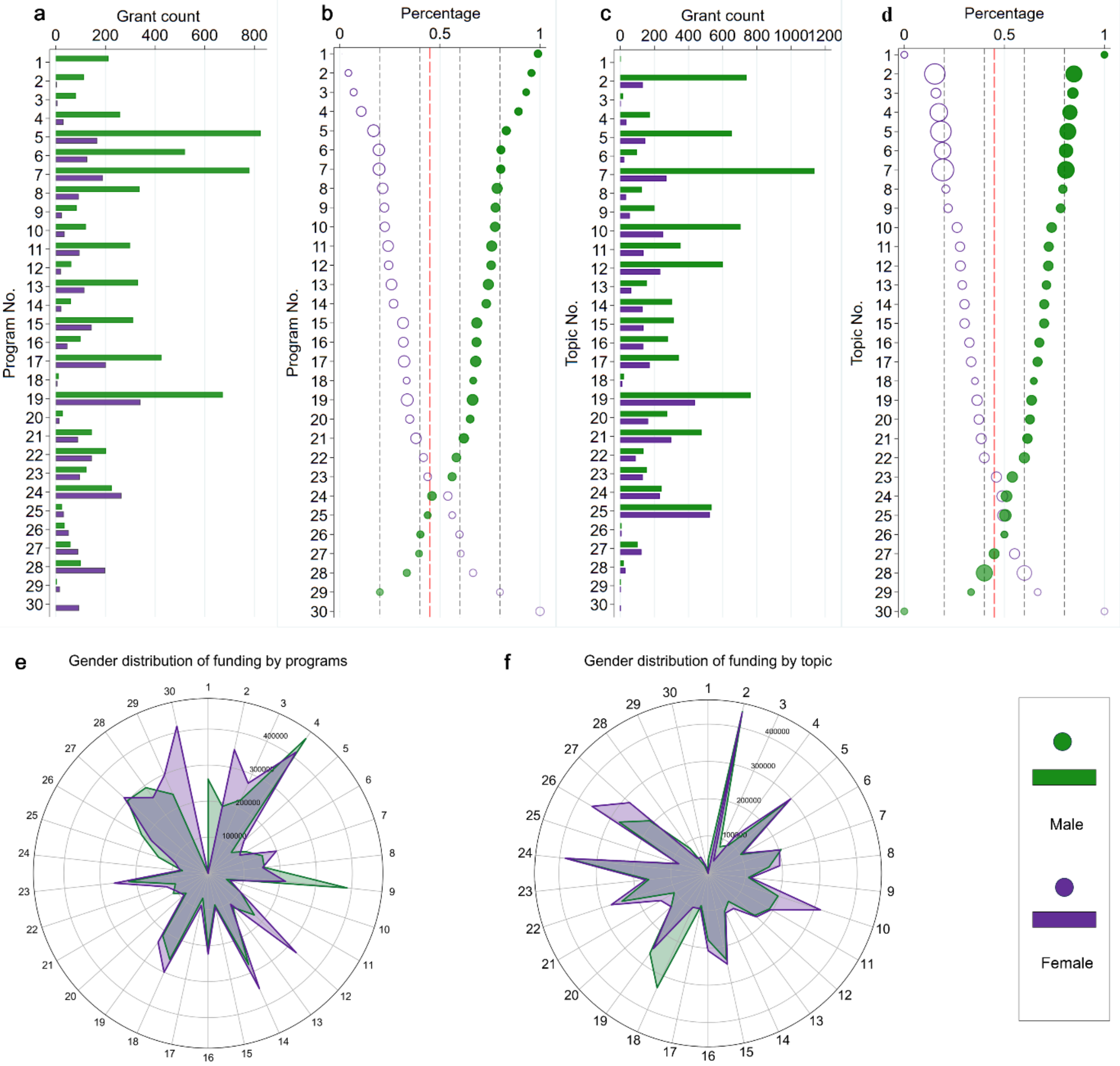

**Fig. 1. Distribution of recipient gender.** The figure illustrates the gender distribution across the top 30 NSF-funded research projects (a, b) and research topics (c, d) in the social sciences, alongside their average funding amounts (e, f). Green represents male scholars, and purple represents female scholars. Panels a and c show the distribution of male and female awardees by project and topic, respectively, while panels b and d depict gender proportions, with the red dashed line indicating the overall gender funding proportion in SBE. Panels e and f present the average funding amounts for male and female scholars in the top 30 projects and topics.

Fig. 1d shows the distribution of recipient gender by research topics. While the number of research topics in which female representation exceeds the overall female funding rate rises to eight, it remains fewer than the number of male-dominated topics. Mirroring the gender imbalance in the official research programs, male recipients are highly represented in topics related to brain, visual, neural, and perceptual research (Topic 25), data and survey-based social analysis (Topic 12), political and democratic studies (Topic 26), and models and methodology (Topic 28), where the proportion of male recipients exceeds 80%. Correspondingly, the proportion of females funded in several technology-, innovation-, and methodology-related topics remains lower than the overall female funding rate. Furthermore, female scholars are relatively more represented in research funding topics related to children and family (Topic 9), health and household risk (Topic 13), learning and cognition (Topic 20), spatial cognition and infants (Topic 15), social and community studies (Topic 29), and health adaptation and pathology (Topic 27).

Female recipients are more concentrated in several topics related to children, family, cognition, and health, while they remain less represented in many other funded topics. Despite claims by the SBE and some scholars that females' funding success in social sciences is comparable to that of males and that gender barriers have been reduced, the topic-level distribution suggests that gender imbalance may persist in a less visible form. Although the number of female applicants may be fewer and their success rate may be comparable to that of males. However, the topic-level distribution indicates that gender imbalance remains visible when we consider the content of the research without considering the official SBE program categories.

**Academic performance of scholars amidst gender-funding imbalance**

Overall, as shown in Fig. 2a, in general research topics, female researchers tend to benefit more from research funding compared to their male counterparts. Female scholars receiving funding experience significantly larger increases in both publication output and citation impact, with gains of 93.4% and 1.08, respectively, compared to 63.1% and 0.57 for men. In terms of journal prestige, measured by CiteScore, the effects are marginal for both genders, with no significant differences observed for either group in these topics [30].

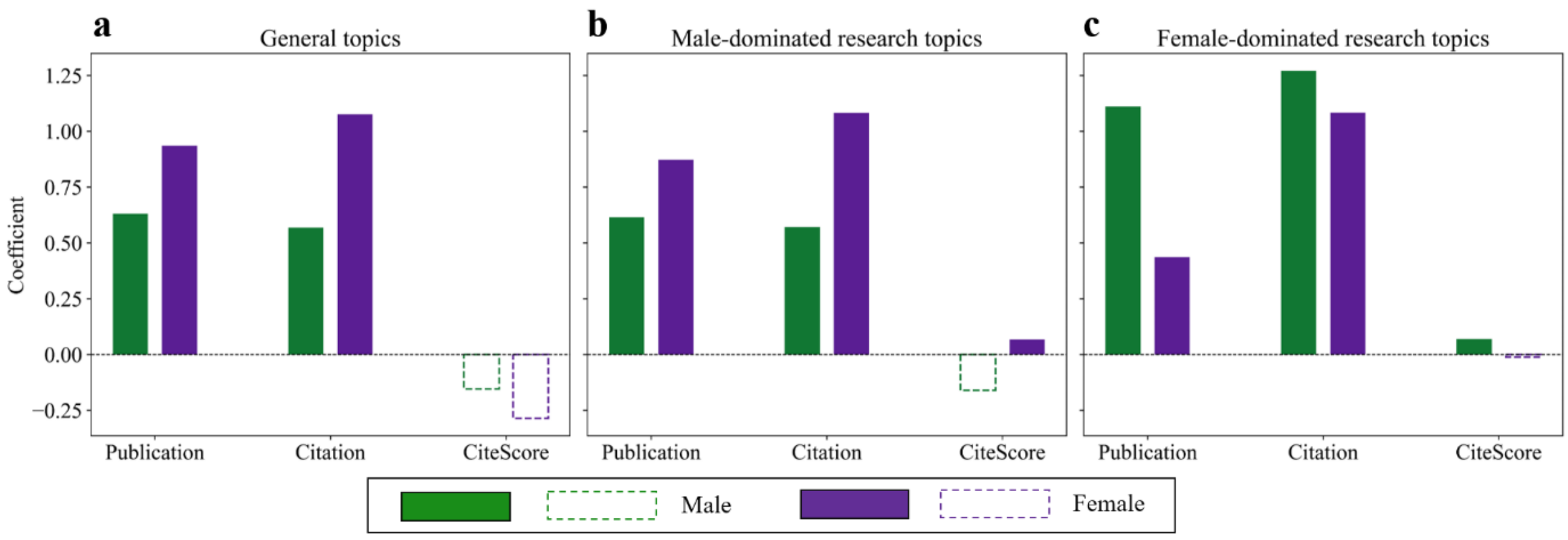

**Fig. 2. Gender differences in the impact of research funding on publications, citations, and CiteScore.**

Dashed bars represent results that are not statistically significant.

A male-dominated research topic is one where the proportion of male recipients exceeds the SBE's average male funding rate, and vice versa for female-dominated topics. In male-dominated research topics (Fig. 2b), the advantage for female researchers persists. Funded women show more significant improvements in both publication output (87.3% vs. 61.5%) and citation counts (108.1% vs. 52.4%) compared to men. Furthermore, women achieve substantial gains in publishing in higher-prestige journals, as indicated by CiteScore, while the effects for men remain relatively negligible. These results suggest that female researchers show larger funding-associated gains in both the quantity and impact of their work in traditionally male-dominated fields. In female-dominated research topics (Fig. 2c), male researchers benefit more from funding. Funded male researchers achieve greater increases in publication output (111.1% vs. 43.6%) and citation counts (1.27 vs. 0.95) compared to female researchers.

However, as seen in Figure 2c, the effect of funding on CiteScore is not significant for female recipients in female-dominated research topics, whereas the estimate for male recipients is positive and marginally significant. These findings imply that male researchers may have more opportunities to maximize the effects of funding in fields traditionally dominated by women, particularly in publication output, citation impact, and journal prestige outcomes. The complete regression results are provided in **Supplementary Note 2**. In **Supplementary Note 3**, we further validated these findings using the propensity score matching to strengthen the robustness of our results.

## Multi-stage estimates of funded scholars' performance

Fig. 3 shows gender-specific patterns of growth or decline of academic performance over post-award time. When disregarding which side of the research topic is dominated (Fig. 3a, 3b, and 3c), we can see that males and females in the social sciences have different publishing efficiency patterns post-award. Male scholars show a decreasing rate of improvement in their publication efficiency. Females' academic performance, in contrast, displays a U-shaped pattern of decline followed by increase, with higher publication efficiency than males in each year.

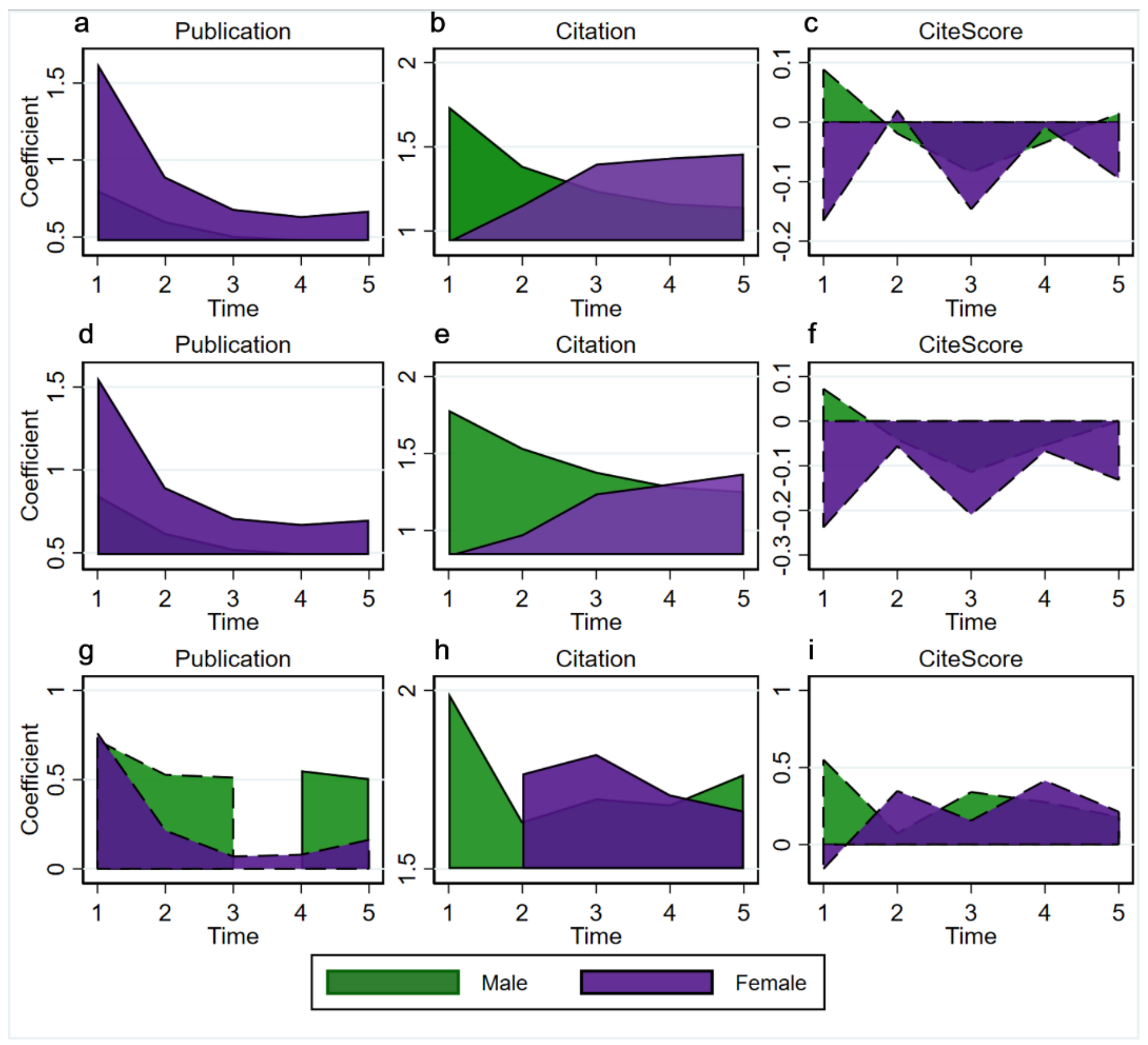

**Fig. 3. Dynamic academic performance of male and female scholars in post-award period.**

Drawing on several studies [31] to understand the diverse manifestations of individual behavior,

it is a typical case that individual differential performance is frequently a combination effect of ability or opportunity and motivation on outcomes. The U-shaped pattern usually means that when scholars receive new treatments (e.g., research grants), they initially begin to establish higher processing strategies and requirements. However, this can result in cognitive overload and a loss of processing efficiency [32]. As scholars become familiar with the new strategies and refine their advanced knowledge, they may have more effective cognition about what they are working on, leading to a slow recovery in productivity [33,34]. Thus, in general, as scholars receive funding to conduct more challenging research, both females and males experience a temporary loss of growing publication efficiency due to strategy reshaping and novel knowledge acquisition. Subsequently, female scholars appear to recover and begin to increase their publication efficiency again; however, while males' publication efficiency remains enhanced by the grant, the continuing deceleration in publication efficiency growth suggests that males may face challenges in sustaining momentum in the short-term, post-award period.

Despite the observed increase in research impact for both males and females after receiving funding, they have completely opposite impact development patterns. As shown in Fig 2(b), the citation improvement rate for male scholars decreases as they move away from the award year, while the citation rate for females shows a gradual increase. In the early post-award period, male scholars are cited more than females, but the situation is reversed from the third post-award year onwards, i.e., the rate of increase in citations for female publications begins to outpace that of males.

Additionally, the pattern of change in the performance for males and females is quite consistent between undifferentiated and male-dominated research topics (Figs. 3d, 3e, and 3f). However, it takes longer ($4^{th}$ year) for females to begin to outperform male scholars in citations for male-dominated research topics (see Fig 3(e)). The slowdown in the growth rate of citations for male scholars in their own dominant research topics is much smoother.

The pattern of change in academic performance for male and female scholars when funded in female-led research topics is elusive. Female recipients show limited funding-associated gains in publication output, while males' publication effectiveness improves significantly from the fourth year onwards, after a period of “silence.” Also, the citation growth rate for females shows a 'silence-up-decline' trend; for male scholars, it is a decline-increase pattern. After a one-year silent period, the citation rate for male scholars eventually overtakes that of females, although female scholars briefly outperform male scholars in the second and fourth years.

**Impact of postdoctoral experience on research performance across genders and research topics**

Postdoctoral experience shows opposite associations with research performance for women and men in male-dominated research funding topics (Fig. 4a). For female researchers, postdoctoral experience is negatively associated with both publication output and citations (coef. = -7.402, $p < 0.001$; coef. = -5.426, $p = 0.004$), while its association with CiteScore is not statistically significant. For male researchers, postdoctoral experience is positively associated with publications and citations (coef. = 1.978 and 0.353, respectively; $p < 0.001$ for both), but not with CiteScore. In female-dominated research funding topics, the pattern is different (Fig. 4b). For female researchers, postdoctoral experience is positively associated with publications and citations (coef. = 0.224, $p = 0.018$; coef. = 0.652, $p < 0.001$), while the CiteScore estimate is not statistically significant. For male researchers, postdoctoral experience is positively associated only with publication output (coef. = 0.351, $p < 0.001$). Its associations with citations and CiteScore are not statistically significant. Thus, in female-dominated topics, postdoctoral experience is consistently linked to higher publication and citation performance for women, whereas for men the significant association appears for publication output.

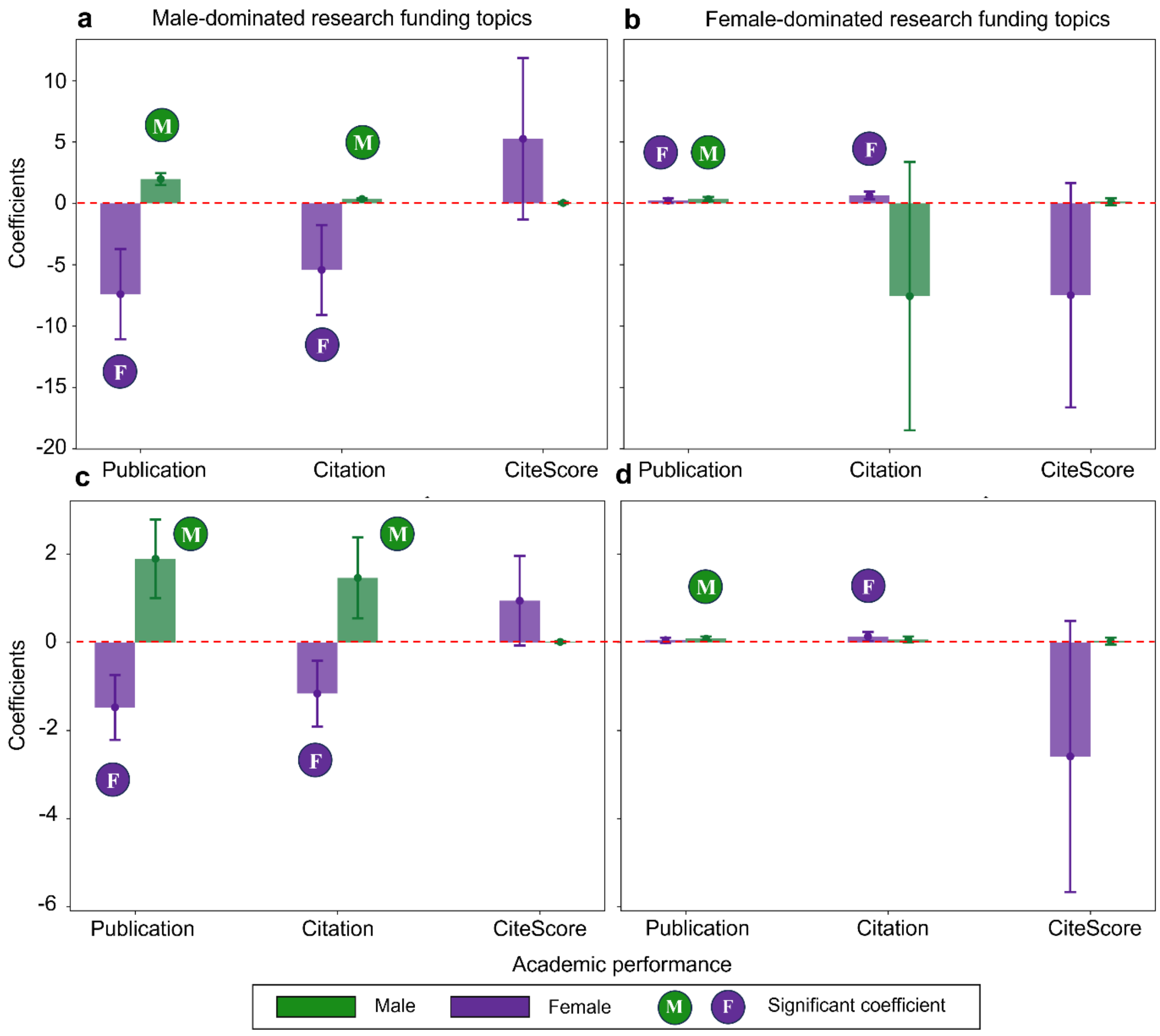

**Fig. 4. Gender differences in the impact of postdoctoral experience on research performance across fields.**

The length of postdoctoral experience produces a similar gender contrast in male-dominated research funding topics (Fig. 4c). For female researchers, longer postdoctoral experience is negatively associated with publications and citations (coef. = -1.473, $p < 0.001$; coef. = -1.161, $p = 0.002$), while the CiteScore estimate is not statistically significant. For male researchers, longer postdoctoral experience is positively associated with both publications and citations (coef. = 1.901, p < 0.001; coef. = 1.467, p = 0.002), but not with CiteScore. In female-dominated research funding topics, longer postdoctoral experience is associated with female researchers' citations, but not with their publication output or CiteScore (Fig. 4d). The citation coefficient is positive and statistically significant (coef. = 0.137, $p$ = 0.009), while the publication and CiteScore estimates are not significant. For male researchers, longer postdoctoral experience is positively associated with publication output (coef. = 0.087, $p < 0.001$), but its associations with citations and CiteScore are not statistically significant.

**DISCUSSION**

The motivation for this study is positive. We focus on one of the heaviest leaks in science pipelines for females in the social sciences; that is, the period from funding to academic performance. We design a "gender -> research funding -> research topic -> academic performance" chain analysis. Using male and female recipients of U.S. social science grants from 2000 to 2019, we first understand the balance in the distribution of research topics when females receive social science grants. Then, to evaluate whether the balance in award topics across gender scholars is reasonable, we use instrumental variables and the 2SLS model to estimate and compare academic performance differences between male and female scholars after receiving grants. Disaggregating the male- and female-dominated research topics, we also provide a dynamic pattern of change in the academic performance for male and female scholars across gender-dominated research topics. We find that the gender distribution of social science funding in research programs and research topics is severely imbalanced. Male scholars remain more represented in most research programs and topics, while there are only a few funding topics that females can dominate. Moreover, social science research topics that are more relevant to technology, information, and mathematics are almost all male-dominated, whilst the female-dominated research topics are related to cognitive research on families, infants, and children. Considering research funding topics tend to be academic frontiers prioritized by the federal government, this also means that academic frontier knowledge in the social sciences is still highly represented by males. This pattern is consistent with topic-level concentration that overlaps with gendered stereotypes in the social sciences: females usually carry an imbalance of family responsibilities and household chores, and then they may be perceived and encouraged to pursue research directions related more to family and household chores as well.

This gender imbalance in funding research topics may further suggest that topic-specific barriers or constraints for female scholars, at least in the early career development and even educational stages prior to funding, are vast and have not been substantially ameliorated. At least, we believe that the current social sciences have been far less effective in correcting the unconscious stereotypes and constraints on females than what Boyle et al. (2015), National Science Foundation (2021), and van der Lee and Ellemers (2015) claim.

We need to acknowledge that grant applications and decisions are often divided by sub-discipline or research project channels rather than research topics. Research topics are frequently interwoven and hidden inside different sub-disciplines or research projects and are not easily perceived by science policymakers. The federal government and funding agencies observe the overall funding probability of females and males on pure numbers tending to equalize. It may mislead reviewers, policy decisionmakers, and funding providers into believing that gender barriers no longer exist in social science. They are less alert to the presence of female stereotypes in the distribution of funding across research content when deciding on grant allocation.

With an unbalanced distribution of genders and research topics, the male and female scholars' academic performances in each other's dominant research topics also show a sharp and dramatic contrast. In both cases, overall or when considering male-dominated research topics, female academics outperform males in terms of publication efficiency and citation growth after receiving funding. In female-dominated research topics, the female advantage disappears and is replaced by males' efficient publication and high-impact performance. Our concern is that, when social science funding and research topics are unevenly distributed by gender, this may be associated with topic sorting and with continued gender differences in early-career trajectories in their early career and even during their education. Further, as funding for most research topics in the social sciences continues to squeeze the space for females, it will naturally also affect the careers of male scholars in female-dominated or even saturated research topics. In other words, allowing barriers to females to persist in funding and research topics will probably leave males who are passionate about female-led research topics to face potential gender barriers as well. Therefore, federal funding agencies could consider topic-level gender representation when evaluating broader participation across research areas. Funding agencies could consider monitoring topic-level gender representation and supporting broader participation across research topics, especially in areas where women remain underrepresented. Broader gender inclusion across both male- and female-dominated research topics may help sustain diversity in social science knowledge production and funding outcomes, although the causal mechanisms behind these patterns require further study.

Furthermore, there are significant differences in the patterns of academic performance of female and male scholars over time after receiving funding. Females tend to show a U-shaped pattern of growth changes in their publication efficiency and increasing rates of research impact, in both overall and male-dominated research topics. The funding contribution to male scholars' research activities in these research topics is decreasing year by year. However, males in social sciences outperform female scholars in both publication efficiency and citations in female-dominated research topics, with larger funding-associated gains. These patterns caution against interpreting concentration in stereotype-related research areas as reflecting scholars' preferences alone, and suggest that scholars can perform strongly in topics not dominated by their own gender when opportunities are available. These findings do not imply that male and female research topics or funding opportunities should be reversed, but it does show that female and male scholars in social sciences can have equally strong research resilience and persistence in research topics that are not dominated by their own gender and numbers.

The results also reveal notable gender differences in how postdoctoral experience relates to later research performance across gender-dominated research funding topics. In male-dominated topics, postdoctoral experience and longer postdoctoral tenure are both associated with lower publication output and citation impact for female researchers, while they are positively associated with publication and citation performance for male researchers. This contrast suggests that postdoctoral training may not operate as the same career resource for women and men in male-dominated research environments. In female-dominated topics, the pattern is less symmetric: postdoctoral experience is associated with higher publication output and citation impact for female researchers, whereas for male researchers the significant association appears mainly in publication output. These findings suggest that early-career training interacts with the gender composition of research topics, reinforcing some inequalities rather than uniformly improving later academic performance.

Finally, we believe that our findings suggest that female scholars in the social sciences may still face topic-specific barriers or constraints. We may have to admit that gender barriers for females may not be resolved in the short term, i.e., it is a long-term and sophisticated endeavor [13]. The gender imbalance in research funding topics, and the flip result of males and females in relation to their academic performance in their dominant research topics, is shaping up to be a dual imbalance. On the one hand, research funding is indeed facilitating males and females in social sciences to boost their vitality. On the other hand, unevenly distributed resources and our evidence suggest that this dual imbalance may be associated with constraints on the research vitality of scholars of both sexes. Whether gender barriers have been reduced in the social sciences is not simply determined by certain statistical indicators for females. We propose that the federal government and funding agencies could pay greater attention to career progression of female scholars in the narrower research content, and examine and address potential sources of gender imbalance in funding allocation and research content.

## ACKNOWLEDGEMENTS

We thank Chao Lu and Lishan Hou for fruitful discussions that helped improve this study. We are also grateful to Julien Larregue for suggesting additional sociological literature that helped strengthen the framing of the manuscript.

## REFERENCES

1. Nickell, D. S. *Guidebook for the Scientific Traveler: Visiting Physics and Chemistry*

*Sites Across America*. (Rutgers University Press, 2010).

2. Freeman, L. A. *Longing for the Bomb: Oak Ridge and Atomic Nostalgia*. (University of North Carolina Press, 2015).

3. Sanderson, K. More women than ever are starting careers in science. *Nature* **596**, 177 (2021).

4. Casad, B. J. *et al.* U.S. women faculty in the social sciences also face gender inequalities. *Front. Psychol.* **13**, (2022).

5. Winslow, S. Gender inequality and time allocations among academic faculty. *Gend. Soc.* **24**, 769–793 (2010).

6. Simpkins, S., Estrella, G., Gaskin, E. & Kloberdanz, E. Latino parents' science beliefs and support of high school students' motivational beliefs: Do the relations vary across gender and familism values? *Soc. Psychol. Educ.* **21**, 1203–1224 (2018).

7. Trusz, S. Why do females choose to study humanities or social sciences, while males prefer technology or science? Some intrapersonal and interpersonal predictors. *Soc. Psychol. Educ.* **23**, 615–639 (2020).

8. Antón, S. C., Malhi, R. S. & Fuentes, A. Race and diversity in U.S. biological anthropology: A decade of AAPA initiatives. *Am. J. Phys. Anthropol.* **165**, 158–180 (2018).

9. Ceci, S. J. & Williams, W. M. Understanding current causes of women's underrepresentation in science. *Proc. Natl. Acad. Sci.* **108**, 3157–3162 (2011).

10. Grant, J., Burden, S. & Breen, G. No evidence of sexism in peer review. *Nature* **390**, 438 (1997).

11. Hutchins, H. M. & Kovach, J. V. ADVANCING women academic faculty in STEM careers: The role of critical HRD in supporting diversity and inclusion. *Adv. Dev. Hum. Resour.* **21**, 72–91 (2019).

12. Boyle *et al.* Gender balance: Women are funded more fairly in social science. *Nature* **525**, 181 (2015).

13. van der Lee, R. & Ellemers, N. Gender contributes to personal research funding success in The Netherlands. *Proc. Natl. Acad. Sci.* **112**, 12349–12353 (2015).

14. National Science Foundation. *National Center for Science and Engineering Statistics*

*[NCSES]. Table 59. Median Annual Salaries of U.S. Residing Full-Time Employed Doctoral Scientists and Engineers in 4-Year Educational Institutions, by Field of Doctorate, Sex, and Faculty Rank: 2019* (2021).

15. Eagly, A. H. Do the social roles that women and men occupy in science allow equal access to publication? *Proc. Natl. Acad. Sci.* **117**, 5553–5555 (2020).

16. Lundberg, S. & Stearns, J. Women in economics: Stalled progress. *J. Econ. Perspect.* **33**, 3–22 (2019).

17. Head, M. G., Fitchett, J. R., Cooke, M. K., Wurie, F. B. & Atun, R. Differences in research funding for women scientists: a systematic comparison of UK investments in global infectious disease research during 1997-2010. *BMJ Open* **3**, e003362 (2013).

18. Burns, K. E. A., Straus, S. E., Liu, K., Rizvi, L. & Guyatt, G. Gender differences in grant and personnel award funding rates at the Canadian Institutes of Health Research based on research content area: A retrospective analysis. *PLoS Med.* **16**, e1002935 (2019).

19. Auriol, E., Friebel, G., Weinberger, A. & Wilhelm, S. Underrepresentation of women in the economics profession more pronounced in the United States compared to heterogeneous Europe. *Proc. Natl. Acad. Sci.* **119**, e2118853119 (2022).

20. Larrègue, J. & Pavie, A. Prestige at play: University hierarchies and the reproduction of funding inequalities. *Can. Rev. Sociol.* **63**, (2025).

21. Schweiger, G. *et al.* The costs of competition in distributing scarce research funds. *Proc. Natl. Acad. Sci.* **121**, e2407644121 (2024).

22. Goulden, M., Mason, M. A. & Frasch, K. Keeping women in the science pipeline. *Ann. Am. Acad. Pol. Soc. Sci.* **638**, 141–162 (2011).

23. Feldon, D. F. *et al.* Postdocs' lab engagement predicts trajectories of PhD students' skill development. *Proc. Natl. Acad. Sci.* **116**, 20910–20916 (2019).

24. Imbens, G. W. & Athey, S. Identification and inference in nonlinear difference-in-difference models. *Econometrica* **74**, 431–497 (2006).

25. De Chaisemartin, C. & D'Haultfoeuille, X. Fuzzy differences-in-differences. *Rev. Econ. Stud.* **85**, 999–1028 (2018).

26. Howell, A. Socio-economic impacts of scaling back a massive payments for ecosystem services programme in China. *Nat. Hum. Behav.* **6**, 1218–1225 (2022).

27. Ding, Y. & Bu, Y. Political hegemony, imitation isomorphism, and project familiarity: Instrumental variables to understand funding impact on scholar performance. *Quant. Sci. Stud.* 1–34 (2025) doi:10.1162/qss_a_00359.

28. Lawson, C., Geuna, A. & Finardi, U. The funding-productivity-gender nexus in science, a multistage analysis. *Res. Policy* **50**, 104182 (2021).

29. Kelchtermans, S., Neicu, D. & Veugelers, R. Off the beaten path: What drives scientists' entry into new fields? *Ind. Corp. Chang.* **31**, 654–680 (2022).

30. Thakuria, A., Deka, D. & Chakraborty, I. Comparative analysis of impact, collaboration, and prestige of open access and subscription based LIS journals using SJR and Scopus indicators. *Int. Inf. Libr. Rev.* **56**, 384–403 (2024).

31. Haans, R. F. J., Pieters, C. & He, Z.-L. Thinking about U: Theorizing and testing U- and inverted U-shaped relationships in strategy research. *Strateg. Manag. J.* **37**, 1177–1195 (2016).

32. Siegler, R. S. Five generalizations about cognitive development. *Am. Psychol.* **38**, 263–277 (1983).

33. Bjorklund, D. F., Miller, P. H., Coyle, T. R. & Slawinski, J. L. Instructing children to use memory strategies: Evidence of utilization deficiencies in memory training studies. *Dev. Rev.* **17**, 411–441 (1997).

34. Pauls, F., Macha, T. & Petermann, F. U-shaped development: An old but unsolved problem. *Front. Psychol.* **4**, (2013).

35. Pell, A. N. Fixing the Leaky Pipeline: Women Scientists in Academia. *J. Anim. Sci.* **74**, 2843–2848 (1996).

36. Sheltzer, J. M. & Smith, J. C. Elite male faculty in the life sciences employ fewer women. *Proc. Natl. Acad. Sci.* **111**, 10107–10112 (2014).

37. Lee, Y. & Won, D. Trailblazing women in academia: Representation of women in senior faculty and the gender gap in junior faculty's salaries in higher educational institutions. *Soc. Sci. J.* **51**, 331–340 (2014).

38. Jackson, R. Women in science: Plugging the leaky pipeline, how can we retain more women in science? *Scientifica* (2013).

39. Moss-Racusin, C. A., Dovidio, J. F., Brescoll, V. L., Graham, M. J. & Handelsman, J. Science faculty's subtle gender biases favor male students. *Proc. Natl. Acad. Sci.* **109**,

16474–16479 (2012).

40. Carvan, T. Why aren't women in science applying for grants? *College of Science, Australian National University* (2022).

41. Bird, D. K. S. Do women publish fewer journal articles than men? Sex differences in publication productivity in the social sciences. *Br. J. Sociol. Educ.* **32**, 921–937 (2011).

42. Ross, M. B. *et al.* Women are credited less in science than men. *Nature* **608**, 135–145 (2022).

43. Malisch, J. L. *et al.* In the wake of COVID-19, academia needs new solutions to ensure gender equity. *Proc. Natl. Acad. Sci.* **117**, 15378–15381 (2020).

44. Buckee, C. *et al.* Women in science are battling both Covid-19 and the patriarchy. *Times higher education* (2020).

45. Currie, J., Thiele, B. & Harris, P. *Gendered Universities in Globalized Economies: Power, Careers, and Sacrifices*. (Lexington Books, 2002).

46. Beaudry, C. & Larivière, V. Which gender gap? Factors affecting researchers' scientific impact in science and medicine. *Res. Policy* **45**, 1790–1817 (2016).

47. Stoop, D., Cobo, A. & Silber, S. Fertility preservation for age-related fertility decline. *Lancet* **384**, 1311–1319 (2014).

48. Madsen, E. B. & Aagaard, K. Concentration of danish research funding on individual researchers and research topics: Patterns and potential drivers. *Quant. Sci. Stud.* **1**, 1159–1181 (2020).

49. King, G. & Nielsen, R. Why Propensity Scores Should Not Be Used for Matching. *Polit. Anal.* **27**, 435–454 (2019).

50. Dong, X. *et al.* Beyond correlation: Towards matching strategy for causal inference in Information Science. *J. Inf. Sci.* 1–14 (2021) doi:10.1177/0165551520979868.

51. Blei, D. M., Ng, A. Y., Jordan, M. I. & Lafferty, J. Latent dirichlet allocation. *J. Mach. Learn. Res.* **3**, 993–1022 (2003).

52. Hausman, J. Specification tests in econometrics. *Econometrica* **46**, 1251–1271 (1978).

53. Lal, A., Lockhart, M., Xu, Y. & Zu, Z. How Much Should We Trust Instrumental Variable Estimates in Political Science? Practical Advice based on Over 60 Replicated

Studies. *Polit. Methods Quant. Methods eJournal* (2021).

54. Murray, M. P. Avoiding Invalid Instruments and Coping with Weak Instruments. *J. Econ. Perspect.* **20**, 111–132 (2006).

55. Bol, T., de Vaan, M. & van de Rijt, A. The Matthew effect in science funding. *Proc. Natl. Acad. Sci. U. S. A.* **115**, 4887–4890 (2018).

56. Antonelli, C. & Crespi, F. The 'Matthew effect' in R&D public subsidies: The Italian evidence. *Technol. Forecast. Soc. Change* **80**, 1523–1534 (2013).

57. Cyrenne, P. & Grant, H. University decision making and prestige: An empirical study. *Econ. Educ. Rev.* **28**, 237–248 (2009).

58. Wedlin, L. The Role of Rankings in Codifying a Business School Template: Classifications, Diffusion and Mediated Isomorphism in Organizational Fields. *Wiley-Blackwell Eur. Manag. Rev.* (2007).

59. Budd, R. Isomorphic tensions and anxiety in UK social science doctoral provision. *Policy Rev. High. Educ.* (2023).

60. Cattaneo, M., Horta, H. & Meoli, M. Dual appointments and research collaborations outside academia: evidence from the European academic population. *Stud. High. Educ.* 2066–2080 (2018) doi:10.1080/03075079.2018.1492534.

61. National Science Board. Chapter 7 science and technology: Public attitudes and understanding. in *Science and Engineering indicators 2018* (2018).

## Supplementary Note 1: Gender, funding topics, and the science pipeline

Pell (1996) was the first to define four developmental stages in the science pipeline for scholars, early childhood, adolescence, college, and graduate school/career. Female scholars are more likely to be lost than male scholars at different stages of the academic and career pipeline, and this is referred to as the 'leaky pipeline' or 'cold climate' [36]. There are many factors that contribute to pipeline leaks for women at the career stage, including gender bias and discrimination, lack of role models and mentoring, and women's unbalanced work-life responsibilities [4]. For example, female scholars are less likely to have access to mentorship and research funding than male scholars, which can limit their opportunities for career advancement. Recently, the federal government and research institutions have been addressing the leakage of women in the science pipeline by increasing the number of leadership positions for women in administrative and research positions, promoting work-life balance, expanding their resources and mentoring opportunities at all stages of the science pipeline, and more [9,37].

The latest evidence suggests that female scholars in the social sciences are experiencing significant improvements in the science pipeline during childhood, adolescence, and college [14]. The biggest leakage crisis facing female scholars in the science pipeline lies in the publication process, grant availability. and promotion of PI at the early-career stage [9,38,39]. Auriol et al. (2022) and Boyle et al. (2015) find that female scholars experience fierce competition after completing their post-doctoral research, failing to break through some invisible barrier. At the early career stage, females often bear disproportionate family responsibilities such as raising children, which leads to them being perceived as not being able to spend more time focusing on their research. The public becomes more distrustful of their scientific accomplishments and leads to a mass exodus of them out of the science pipeline in the fourth stage [28,40–42].

This gender barrier was particularly prominent during the COVID-19 pandemic [43]:35 female scientists from around the world protested that they were experiencing severe gender discrimination. Female research outputs have a low to zero citation rate, with damaging effects on female professional and academic development [44]. Thus, while the award structure in academia tends to value typical attributes and characteristics of men [45], it seems that there is only an incentive to embrace women in some low-impact research topics.

Research investigating the leakage of female scholars in the fourth stage of the science pipeline is not uncommon. Existing research mostly focuses on whether female scholars at a single subject or multidisciplinary level are equitable in the funding distribution [46], i.e., funding-gender relations research. Alternatively, some scholars justify the irreplaceable position of female scholars in scientific research by highlighting their scientific output once they enter their career [47], outlined as gender-productivity relations research. However, the focus on the mere increase in the number or success of female scholars in research grants and the lack of assessment of post-funding research feedback from female scholars seem somewhat blinkered. Gender stratification of academic performance while ignoring policy interventions (e.g., research funding) also does not seem to provide straightforward policy advice to policymakers. Thus, recent evidence suggests that some scholars are beginning to take note of the 'gender-research funding and academic performance' chain, providing information on the gender gap in the academic productivity of grantees after funding [28].

However, ignoring the research content to compare the representation of male and female scholars may cause female scholars to struggle with 'stagnation' in certain research topics, particularly male-dominated research topics. Further, we are still unable to understand the academic performance of funded male and female scholars across different research topics, which does not provide a more compelling evidence base for rationalizing the allocation of academic resources among research topics. The untargeted encouragement of female scholars and legislation may even raise concerns about male inequality [13]. Therefore, it is imperative to establish a chain of gender, research funding, research topics, and academic performance for female scholars in social sciences, as illustrated in the Supplementary Figure 1. We need evidence to further understand the academic performance of female scholars in every funded research topic, to prevent gender gaps in academic performance caused by over- and under-funding of women in certain research topics, and to protect female scholars from leaking out of the science pipeline again [15,48].

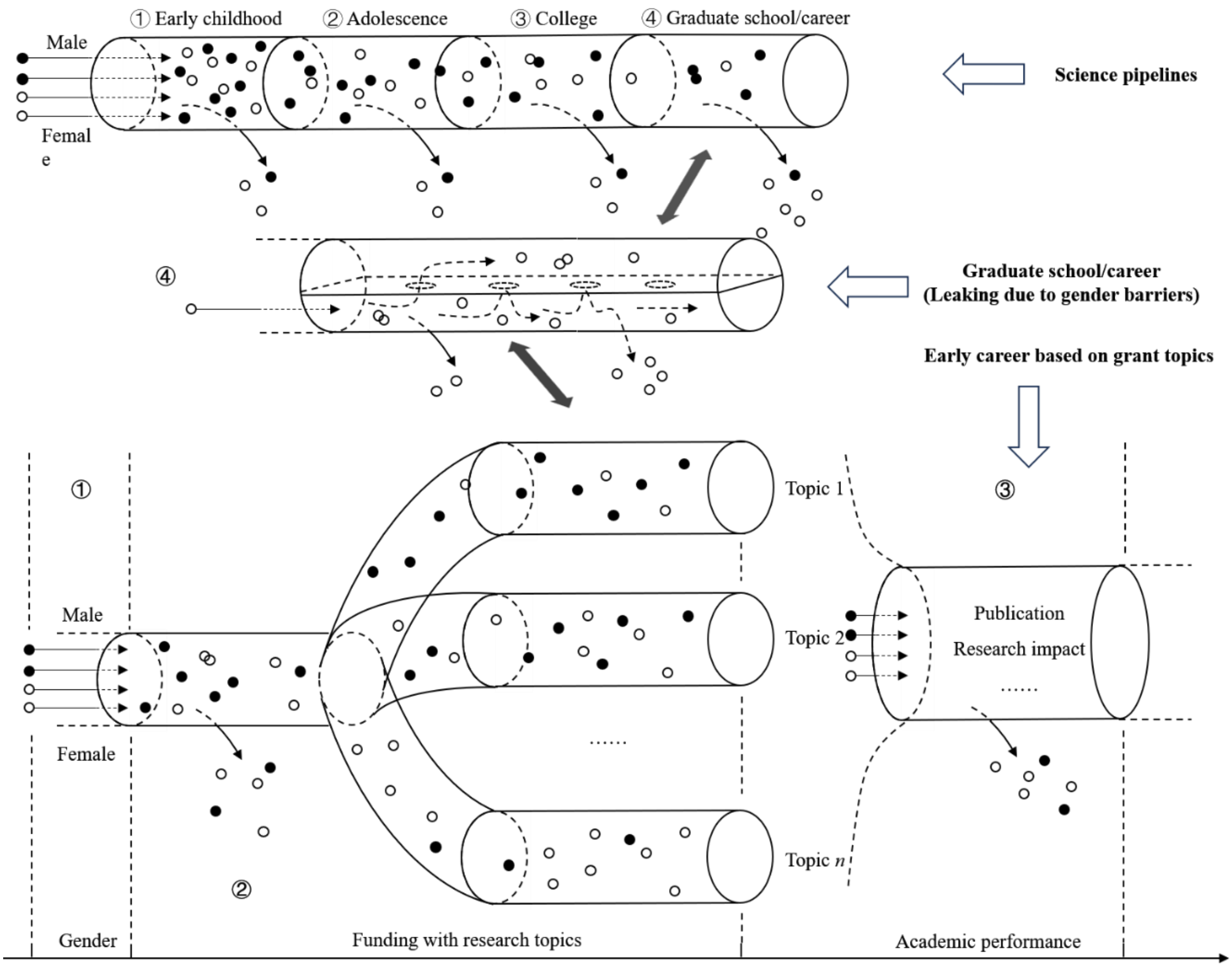

**Supplementary Figure 1. Gender, research funding topics, grants and scientific productivity.**

## Supplementary Note 2: The regression results for the research funding effect on scholars' academic performance.

Supplementary Table 1 shows the academic performance of male and female scholars after receiving a social science grant. Research funding in social sciences has a significant positive effect on scholars' academic performance, and female scholars outperform male scholars. The average publication efficiency of males is improved by 63.05% following the stimulation of research funding, while female scholars publish approximately 30% more efficiently than males, up to 93.44%. Female publications' impact increased (coef=1.0758), again more than males' impact (coef=0.5682). Yet, funding does not lead to male and female scholars' research being published in more prestigious journals. These results imply the existence of the Calutron Girl phenomenon in social sciences, whereby females have the opportunities to outperform males in terms of productivity and impact. Improvements in publication productivity and research impact are usually associated with scholars' desire to upgrade their research resources, establish priority in their research direction, and mitigate the dilemma of suppression due to gender (Fox, 2005; Mullins, 1973). Thus, this may imply that female scholars in social sciences are more proficient in using research grants. Also, research funding and academic performance beyond that of male scholars will likely allow female scholars to further reduce the productivity gap that may be caused by gender.

**Supplementary Table 1. Evaluations of academic performance for males and females after receiving funding.**

| Variable | Male | | | Female | | |
|---|---|---|---|---|---|---|
| | Article | Citation | CiteScore | Article | Citation | CiteScore |
| Funding | 0.6305***<br>(0.0648) | 0.5682***<br>(0.0945) | -0.1536<br>(0.1206) | 0.9344***<br>(0.1349) | 1.0758***<br>(0.0271) | -0.2855<br>(0.2294) |
| Initial performance | 0.1501***<br>(0.0133) | 0.3774***<br>(0.0206) | -0.0134<br>(0.0251) | 0.1928***<br>(0.0255) | 0.5057***<br>(0.0081) | -0.0507***<br>(0.0449) |
| Funding quota | 0.0079*<br>(0.0032) | -0.0092*<br>(0.0043) | 0.0162**<br>(0.0049) | 0.0322***<br>(0.0049) | 0.0058<br>(0.0096) | 0.0267**<br>(0.007) |
| Career age | 0.0167***<br>(0.0008) | -0.0030***<br>(0.0006) | -0.0019**<br>(0.0007) | 0.0205***<br>(0.0017) | -0.0056***<br>(0.0014) | -0.004***<br>(0.0013) |
| Award year | 0.0256***<br>(0.0014) | -0.0790***<br>(0.0023) | -0.0084***<br>(0.0022) | 0.0368***<br>(0.0027) | -0.0674***<br>(0.0031) | -0.0195***<br>(0.0049) |
| Affiliations' articles | -0.0561*** | -0.2886*** | -0.2307*** | -0.0277*** | -0.1775*** | -0.2715*** |
| | (0.0072) | (0.0101) | (0.0126) | (0.0098) | (0.0141) | (0.0175) |
| Fields' articles | 0.0249* | -0.2807*** | -0.3874*** | 0.0715** | -0.2057*** | -0.3583*** |
| | (0.0124) | (0.0174) | (0.0206) | (0.0200) | (0.0285) | (0.0300) |
| Fields' citations | 0.0622*** | 0.2995*** | 0.2438*** | 0.0346*** | 0.1816*** | 0.2844*** |
| | (0.0075) | (0.0106) | (0.0132) | (0.0103) | (0.0148) | (0.0182) |
| Affiliations' citations | -0.0224* | 0.2700*** | 0.3774*** | -0.0628** | 0.1860*** | 0.3506*** |
| | (0.0113) | (0.0158) | (0.0194) | (0.0174) | (0.0251) | (0.0271) |
| Employer reputation | 0.0004* | 0.0010*** | 0.0004 | 0.0003 | 0.0014** | 0.0003 |
| | (0.0002) | (0.0003) | (0.0003) | (0.0000) | (0.0005) | (0.0004) |
| USNEWS ranking | 0.0001*** | 0.0001*** | 0.0000 | 0.0001** | 0.0000 | 0.0001 |
| | (0.0000) | (0.0000) | (0.0000) | (0.0000) | (0.0000) | (0.0000) |
| QS ranking | 0.0000** | 0.0000 | 0.0000 | 0.0001 | 0.0000 | 0.0000 |
| | (0.0000) | (0.0000) | (0.0000) | (0.0000) | (0.0000) | (0.0000) |
| _cons | -51.5325*** | 159.1585*** | 17.9037*** | -74.8003*** | 136.0993*** | 40.1395*** |
| | (2.7415) | (4.7293) | (4.478) | (5.6015) | (6.1817) | (10.0248) |

Note: Initial performance indicates the initial articles or citations or CiteScore of scholars.

Standard errors in parentheses

* p<0.05 ** p<0.01 *** p<0.001

In male-dominated research topics, female scholars who receive funding exhibit higher publication efficiency, citation counts, and journal prestige compared to their male peers. As shown in Supplementary Table 2, the gap between the improvement in publication efficiency of female and male scholars narrowed slightly. For example, the publication efficiency of female scholars is raised by more than 87.25%, far exceeding the 61.50% of male scholars. The citation count of research outcomes published by females on male-dominated research topics is significantly increased by 108.14%, which is even higher than the female performance (coef=1.0758), regardless of the classification of the research topic (see Supplementary Table 2). More importantly, females who are funded in male-dominated research topics are able to publish their research in more prestigious journals; whereas funding in male-dominated research topics has no significant effect on enhancing the prestige of publications for males.

**Supplementary Table 2. Academic performance of funded scholars in male-dominated funding topics.**

| Variable | Male | | | Female | | |
|---|---|---|---|---|---|---|
| | Article | Citation | CiteScore | Article | Citation | CiteScore |
| Funding | 0.6150*** | 0.5702*** | -0.1602 | 0.8725*** | 1.0814*** | 0.0676*** |
| | (0.0655) | (0.0951) | (0.1218) | (0.1402) | (0.0299) | (0.0145) |
| Initial performance | 0.1471*** | 0.3779*** | -0.0156 | 0.1903*** | 0.5071*** | 0.011** |
| | (0.0134) | (0.0208) | (0.0254) | (0.0268) | (0.0089) | (0.0042) |
| Funding quota | 0.0067* | -0.0145** | 0.0112* | 0.0285*** | 0.0141 | 0.0086 |
| | (0.0034) | (0.0045) | (0.0052) | (0.0054) | (0.0109) | (0.0046) |
| Career age | 0.0162*** | -0.0029*** | -0.0020** | 0.0190*** | -0.0063*** | -0.0008 |
| | (0.0008) | (0.0006) | (0.0007) | (0.0018) | (0.0015) | (0.0007) |
| Award year | 0.0256*** | -0.0794*** | -0.0088*** | 0.0384*** | -0.0663*** | -0.0038* |
| | (0.0014) | (0.0024) | (0.0023) | (0.003) | (0.0034) | (0.0015) |
| Affiliations' articles | -0.0565*** | -0.2862*** | -0.2314*** | -0.0258* | -0.1760*** | -0.0586*** |
| | (0.0075) | (0.0105) | (0.0132) | (0.0108) | (0.0155) | (0.0077) |
| Fields' articles | 0.0208 | -0.2748*** | -0.3885*** | 0.0679** | -0.1918*** | -0.0826*** |
| | (0.0127) | (0.0178) | (0.0212) | (0.0232) | (0.0309) | (0.0153) |
| Fields' citations | 0.0631*** | 0.2968*** | 0.2439*** | 0.0324** | 0.1797*** | 0.0617*** |
| | (0.0078) | (0.011) | (0.0139) | (0.0113) | (0.0163) | (0.0081) |
| Affiliations' citations | -0.0180 | 0.2658*** | 0.3807*** | -0.0602** | 0.1711*** | 0.0849*** |
| | (0.0115) | (0.0161) | (0.0198) | (0.0200) | (0.0273) | (0.0136) |
| Employer reputation | 0.0003 | 0.0012*** | 0.0004 | 0.0004 | 0.0017** | -0.0001 |
| | (0.0002) | (0.0003) | (0.0003) | (0.0003) | (0.0005) | (0.0003) |
| USNEWS ranking | 0.0001*** | 0.0001*** | 0.0001 | 0.0000* | 0.0000 | 0.0001 |
| | (0.0000) | (0.0000) | (0.0000) | (0.0000) | (0.0000) | (0.0000) |
| QS ranking | 0.0000*** | 0.0001* | 0.0000 | 0.0000 | 0.0000 | 0.0000* |
| | (0.0000) | (0.0000) | (0.0000) | (0.0000) | (0.0000) | (0.0000) |
| _cons | -51.5310*** | 160.0643*** | 18.8907*** | -77.7163*** | 133.8664*** | 8.0603*** |
| | (2.8520) | (4.8919) | (4.6291) | (6.1513) | (6.8014) | (3.0175) |

Note: Initial performance indicates the initial articles or citations or CiteScore of scholars.

Standard errors in parentheses

* p<0.05 ** p<0.01 *** p<0.001

On the one hand, females in the social sciences may be working much harder than the public believes, not only outperforming male scholars, but also accelerating knowledge discovery in male-dominated research topics. On the other hand, social science funding can change the dilemma that Chatterjee and Werner (2021), Dion et al. (2018) and Lloreda (2022) have found: that female research findings are less likely to be cited by scholars in male-dominated research fields.

Males outperformed females in female-dominated research topics after receiving funding. As shown in Supplementary Table 3, the publication efficiency of female recipients in female-dominated research topics is improved by 43.57 %, which is less than half of the publication efficiency of female scholars in male-dominated topics. However, male publication efficiency is improved by 111.12%, almost double twice the efficiency of their male peers from male-dominated research topics. Also, male scholars who received funding in female-led research show a higher research impact than female scholars, with an improvement of 127.10 %, which surpasses the increase of female scholars' research impact. Dramatically, male scholars being funded in female-dominated research topics enabled male scholars to publish their research in more prestigious journals, with CiteScore being boosted by 6.97 %. Females in their own-dominated topics are unable to publish their research in more prestigious journals.

**Supplementary Table 3. Academic performance of funded scholars in female-dominated funding topics.**

| | Male | | | Female | | |
|---|---|---|---|---|---|---|
| | **Article** | **Citation** | **CiteScore** | **Article** | **Citation** | **CiteScore** |
| Funding | 1.1112*** | 1.2710*** | 0.0697* | 0.4357* | 1.0832*** | -0.0116 |
| | (0.2696) | (0.0704) | (0.0338) | (0.2226) | (0.0649) | (0.0295) |
| Initial performance | 0.2470*** | 0.5387*** | 0.0242* | 0.0775 | 0.5041*** | -0.0035 |
| | (0.0568) | (0.0209) | (0.0099) | (0.0424) | (0.0190) | (0.0085) |
| Funding quota | 0.0258* | 0.0278 | 0.0104 | 0.0173 | -0.0450 | 0.0038 |
| | (0.0103) | (0.0232) | (0.0093) | (0.0096) | (0.0234) | (0.0094) |
| Career age | 0.0237*** | -0.0041 | 0.0003 | 0.0217*** | -0.0016 | 0.0014 |
| | (0.0016) | (0.0030) | (0.0012) | (0.0024) | (0.0037) | (0.0016) |
| Award year | 0.0290*** | -0.0527*** | -0.0021 | 0.0253*** | -0.0741*** | -0.0001 |
| | (0.0045) | (0.0075) | (0.0031) | (0.0044) | (0.0074) | (0.0031) |
| Affiliations' articles | -0.0394 | -0.1948*** | -0.0462** | -0.0597** | -0.1834*** | -0.1079*** |
| | (0.0262) | (0.0370) | (0.0180) | (0.0221) | (0.0344) | (0.0157) |
| Fields' articles | 0.0358 | -0.3012** | -0.1159** | 0.0940* | -0.3128*** | -0.0827* |
| | (0.0590) | (0.0923) | (0.0405) | (0.0426) | (0.0753) | (0.0364) |
| Fields' citations | 0.0400 | 0.2103*** | 0.0544** | 0.0650** | 0.1884*** | 0.1173*** |
| | (0.0274) | (0.0390) | (0.0189) | (0.0234) | (0.0361) | (0.0165) |
| Affiliations' citations | -0.0440 | 0.2807** | 0.1162** | -0.0932** | 0.2936*** | 0.0998** |
| | (0.0534) | (0.0851) | (0.0375) | (0.0364) | (0.0666) | (0.0326) |
| Employer reputation | 0.0030*** | -0.0023 | 0.0001 | 0.0007 | -0.0001 | -0.0001 |
| | (0.0007) | (0.0012) | (0.0005) | (0.0006) | (0.0012) | (0.0005) |
| USNEWS ranking | 0.0001 | 0.0001 | 0.0000 | 0.0001 | 0.0001 | 0.0000 |
| | (0.0001) | (0.0001) | (0.0000) | (0.0000) | (0.0001) | (0.0000) |
| QS ranking | 0.0000 | 0.0000 | 0.0000 | 0.0000 | 0.0000 | 0.0000 |
| | (0.0001) | (0.0001) | (0.0000) | (0.0000) | (0.0001) | (0.0000) |
| _cons | -58.9533*** | 105.5325*** | 4.3139 | -50.8508*** | 149.5943*** | 0.2196 |
| | (9.068) | (15.0332) | (6.1494) | (8.8579) | (14.8517) | (6.193) |

Note: Initial performance indicates the initial articles or citations or CiteScore of scholars.

Standard errors in parentheses

* p<0.05 ** p<0.01 *** p<0.001

The benefits of being a female scholar seem to disappear when it comes to female-dominated research topics (Muench et al., 2014). As some scholars reluctantly remark: “What's the best way to succeed in a female-dominated career? --Be a man.” (Hu, 2015). When federal funding for these family-, baby-, and child-related research topics is allocated to females who are more likely to be exposed to them, their research performance lags behind that of males. Conversely, male scholars' publication effectiveness and research impact on these research topics not only outperforms that of females in the same research topic, but also surpasses the performance of males who lead their own research topics.

In conditions of significant gender imbalance in both research programs and research topics, female and male scholars are showing contrasting academic performance in their respective dominant research content. This is reinforcing the serious consequences of the funding imbalance. On the one hand, the overall funding probability of the social sciences masks gender barriers within the research content, causing female scholars to struggle with leaky pipelines. On the other hand, females and males outperform each other academically in research topics dominated by the opposing side. The damage of this dual imbalance may be incalculable in that scholars who may be more adept at each other's dominant research topics continue to struggle to break through gender barriers into more research topics, thereby affecting the research vitality of social sciences.

## Supplementary Note 3: Propensity score matching (PSM) validation and robustness check

In this study, we used Propensity Score Matching (PSM) as a robustness check to validate the main results obtained using the Instrumental Variables (IV) and Two-Stage Least Squares (2SLS) method presented in the main text. One of the key challenges in applying PSM lies in the panel structure of the original dataset. Since researchers' academic output evolves dynamically over time, directly performing matching on panel data may lead to suboptimal matching quality [49]. To address this issue, we first transformed the dataset to make it more suitable for PSM estimation.

To ensure the effectiveness of the PSM approach, we computed the pre-funding and post-funding averages of annual publication counts and citation counts for each researcher. This transformation resulted in two observations per researcher: one representing the pre-funding period and the other representing the post-funding period. This aggregation removes time dependence within individual trajectories, mitigating potential biases in the matching process [50].

For researchers who had never received funding, we randomly assigned a pseudo research grant within their unfunded period to create a pseudo pre-funding and pseudo post-funding period. We ensured that the duration of the pseudo pre- and post-funding periods matched the length of the actual pre- and post-funding periods for funded researchers (i.e., both spanning five years before and after funding). This approach allowed us to construct a well-defined control group for PSM [50]. Therefore, the treatment group in the PSM framework consisted of researchers who actually received funding, while the control group comprised researchers who were artificially assigned a pseudo funding period.

The propensity scores were estimated using a logistic regression model that incorporated all covariates from the first stage of the 2SLS estimation, along with the instrumental variable and pre-funding academic indicators (following the approach of Ding, 2019). The logistic model was specified as follows:

$$P\left(Treatment=1\,|\,X\right)=\frac{1}{1+e^{-\left(\omega_0+\omega_1X_1+\omega_2X_2+\ldots+\omega_kX_k\right)}} \quad (3)$$

Here, $X_1, X_2, \ldots, X_k$ represent the covariates (all covariates from the first stage regression of 2SLS, along with the instrumental variables), and $\omega_0, \omega_1, \ldots, \omega_k$ are the coefficients to be estimated. Propensity scores were calculated for each scholar, which allowed for matching treated and control groups based on their likelihood of receiving funding.

We performed nearest neighbor matching (NNM) with a caliper that was dynamically adjusted based on the effectiveness of standardized bias (SB) reduction, ensuring an optimal balance between matching quality and sample retention. The caliper selection process followed an iterative approach, where different caliper values were tested, and the final threshold was chosen to minimize the standardized bias while maintaining a sufficient number of matched pairs. The standardized bias for each covariate was calculated using the formula:

$$SB\left(X\right)=\frac{\left|\bar{X}_{treated}-\bar{X}_{control}\right|}{\sqrt{\frac{s_{treated}^2-s_{control}^2}{2}}} \quad (4)$$

where $\bar{X}_{treated}$ and $\bar{X}_{control}$ represent the means of the covariate for the treated and control groups, respectively, and $s_{treated}^2$ , $s_{control}^2$ are the variances in the treated and control groups.

The analysis of matching effectiveness reveals that the matching procedure significantly enhances covariate balance. Prior to matching, notable imbalances were observed in key variables such as institutional reputation, academic ranking, and field citations (see Supplementary Figure 2). Post-matching, the standardized bias for most covariates was substantially reduced, with values converging near zero. These results underscore the efficacy of the matching process in mitigating covariate discrepancies, ensuring a more balanced and robust comparison between the matched samples across both male- and female-dominated fields.

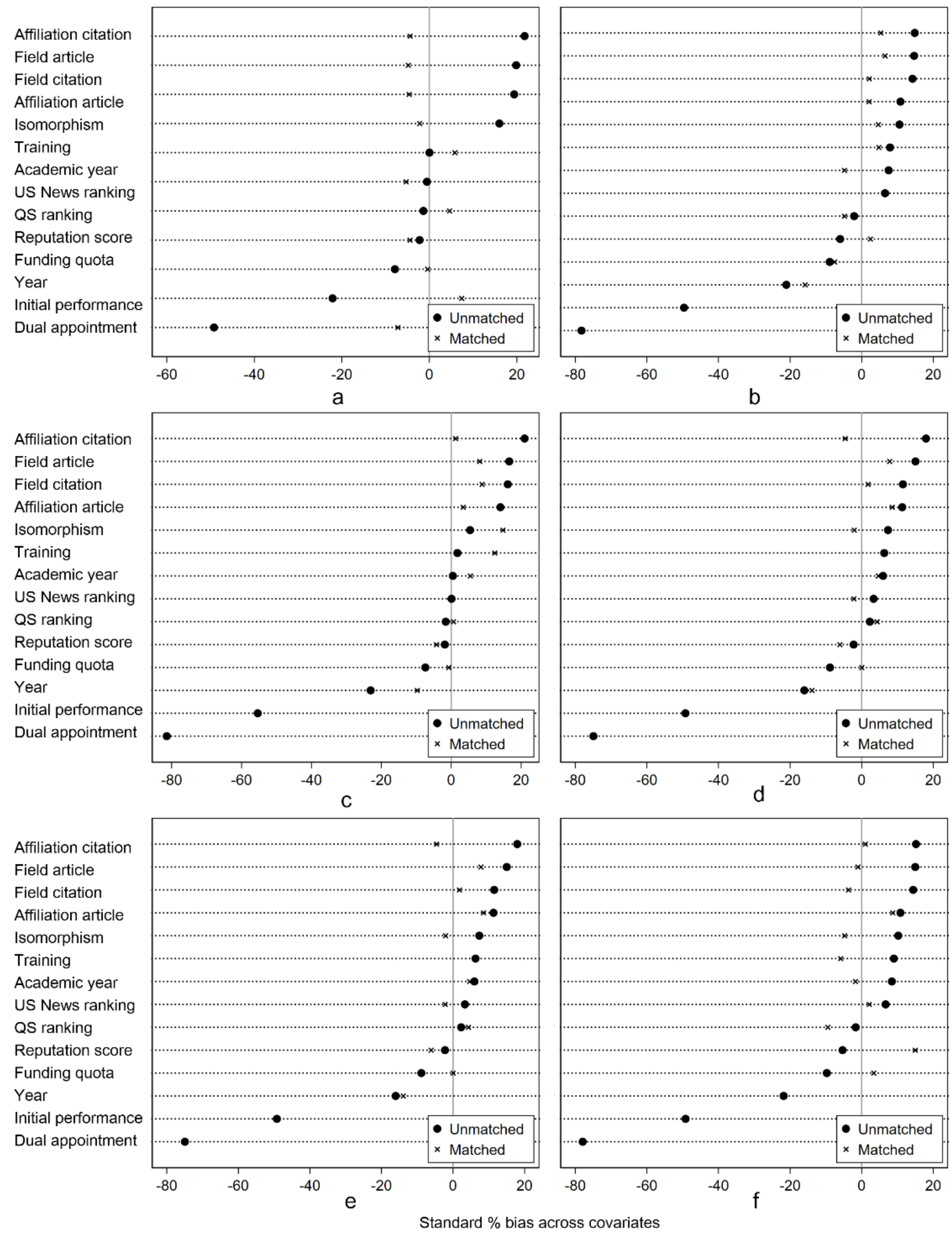

**Supplementary Figure 2. Standardized bias before and after matching in PSM analysis.** This figure illustrates the standardized percentage bias across covariates before and after propensity score matching (PSM)

for various research topics and gender groups. Subfigure (a) and (b) show the overall research topics, with results for male (a) and female (b) scholars. Subfigure (c) and (d) depict male-dominated research fields, highlighting the reduction in bias for male (c) and female (d) researchers. Subfigure (e) and (f) present female-dominated research fields, displaying the results for male (e) and female (f) researchers.

After matching (see Supplementary Table 4), we estimated the treatment effects of research funding on academic performance, including publication output and citation counts. The results of the matching analysis are consistent with those obtained from the 2SLS method. In general research topics, the ATT estimates show that female scholars experienced slightly larger gains than male scholars in both publication output (0.108 vs. 0.091) and citation impact (0.639 vs. 0.621).

**Supplementary Table 4. PSM results.**

| Scholar | Topic | Academic performance | Sample | Treated | Controls | Difference | S.E. | T-stats |
|---|---|---|---|---|---|---|---|---|
| Male | Overall | Publication | Unmatched | 0.676 | 0.552 | 0.124 | 0.020 | 6.210 |
| | | | ATT | 0.676 | 0.585 | 0.091 | 0.050 | 1.830 |
| | | Citation | Unmatched | 1.926 | 1.972 | -0.045 | 0.041 | 1.110 |
| | | | ATT | 1.855 | 1.234 | 0.621 | 0.121 | 5.140 |
| | Male-dominated | Publication | Unmatched | 0.666 | 0.579 | 0.087 | 0.020 | 4.280 |
| | | | ATT | 0.601 | 0.498 | 0.103 | 0.051 | 2.000 |
| | | Citation | Unmatched | 1.918 | 1.958 | -0.041 | 0.045 | 0.910 |
| | | | ATT | 1.818 | 1.365 | 0.454 | 0.113 | 4.030 |
| | Female-dominated | Publication | Unmatched | 0.641 | 0.575 | 0.067 | 0.044 | 1.520 |
| | | | ATT | 0.631 | 0.211 | 0.420 | 0.185 | 2.280 |
| | | Citation | Unmatched | 1.841 | 1.855 | -0.014 | 0.092 | 0.150 |
| | | | ATT | 1.867 | 0.692 | 1.175 | 0.360 | 3.260 |
| Female | Overall | Publication | Unmatched | 0.565 | 0.479 | 0.085 | 0.024 | 3.520 |
| | | | ATT | 0.565 | 0.457 | 0.108 | 0.047 | 2.300 |
| | | Citation | Unmatched | 1.710 | 1.743 | -0.033 | 0.054 | 0.610 |
| | | | ATT | 1.710 | 1.071 | 0.639 | 0.143 | 4.470 |
| | Male-dominated | Publication | Unmatched | 0.549 | 0.516 | 0.033 | 0.025 | 1.310 |
| | | | ATT | 0.554 | 0.446 | 0.107 | 0.050 | 2.140 |
| | | Citation | Unmatched | 1.707 | 1.789 | -0.082 | 0.060 | 1.370 |
| | | | ATT | 1.707 | 1.168 | 0.539 | 0.184 | 2.920 |
| | Female-dominated | Publication | Unmatched | 0.565 | 0.407 | 0.158 | 0.037 | 4.290 |
| | | | ATT | 0.565 | 0.384 | 0.180 | 0.090 | 2.000 |
| | | Citation | Unmatched | 1.700 | 1.559 | 0.141 | 0.105 | 1.350 |
| | | | ATT | 1.700 | 1.153 | 0.546 | 0.245 | 2.230 |

In male-dominated research topics, female scholars also showed slightly larger ATT estimates than male scholars in both publication output (0.107 vs. 0.103) and citation impact (0.539 vs. 0.454). These findings suggest that female researchers are able to effectively leverage research funding to improve both the quantity and impact of their work in male-dominated fields.

In contrast, in female-dominated research topics, male scholars benefitted more from funding, showing larger ATT estimates than female scholars in publication output (0.420 vs. 0.180) and citation impact (1.175 vs. 0.546). This suggests that male researchers may be better positioned to maximize the impact of funding in traditionally female-dominated fields.

Overall, the results from the PSM robustness check align with those obtained using the 2SLS method, suggesting that the main directional patterns are robust after accounting for observable pre-treatment differences through matching. The balance checks, coupled with the matching analysis, confirm that research funding has a differential association with male and female scholars' academic performance, with females generally benefiting more in male-dominated fields and males benefiting more in female-dominated fields.

## Supplementary Note 4: Research topics for grant proposals in the social sciences

Latent Dirichlet Allocation (LDA) is an unsupervised machine learning algorithm used for topic modeling, commonly employed to extract thematic information from textual data [51]. In our study, we utilized the LDA algorithm to extract research topics from funding proposals in the social sciences, aiming to better understand the differences in various research topics and the distribution of male and female scholars within them. We collected data on NSF's SBE-funded projects, comprising 12,945 research proposals. Subsequently, we preprocessed the textual descriptions of these projects, which included tasks such as removing stop words, punctuation, and numerical values, as well as performing stemming and lemmatization to ensure data accuracy and consistency. Then, we transformed the preprocessed textual data into a bag-of-words model, representing each funded project as a term frequency vector that encapsulated the frequency of occurrence of each word in the text. These bag-of-words vectors were then utilized as input data to apply the LDA algorithm for extracting research topics from the projects.

The process of calculating keywords or representative terms for each topic involves estimating the probability distribution of words given topics. This is based on the underlying assumption that each research proposal is generated from a mixture of topics, and each topic is characterized by a distribution over words. The equation to compute the probability of a word, $w$, given a topic, $z$, is,

$$p(w|z) = \frac{n_{w,z} + \kappa}{\sum_{w'} (n_{w',z} + \kappa)} \tag{5}$$

Where $n_{w,z}$ is the number of times word $w$ is assigned to topic $z$ across all proposals. $\kappa$ is a smoothing parameter. The denominator is the sum of the counts of all words in the vocabulary, plus $\kappa$ times the vocabulary size. This ensures that the probabilities sum up to 1. This formula calculates the likelihood of observing word $w$ given topic $z$. It measures how likely it is for a particular word to be associated with a specific topic. Once these probabilities are estimated for each word-topic pair, the top $k$ words with the highest probabilities for each topic are selected as the representative keywords. These keywords provide insights into the main themes or subjects associated with each topic, aiding in the interpretation and understanding of the topics generated. In this study, we assigned all research proposals to 30 research topics, each represented by five keywords, as shown in Supplementary Table 5.

**Supplementary Table 5. Number of topics and keywords for SBE grants.**

| Topic ID | Keywords | Percentage |
|---|---|---|
| 1 | sign, deaf, varieties, heat, asl | 0.42% |
| 2 | brain, visual, neural, human, perception | 6.30% |
| 3 | religious, history, immigration, aids, region | 0.43% |
| 4 | women, gender, race, labor, race | 1.97% |
| 5 | data, survey, social, analysis, time | 6.19% |
| 6 | political, party, democratic, politics, parties | 1.52% |
| 7 | models, theory, methodology, analysis, develop | 8.38% |
| 8 | public, media, attitude, survey, policy | 1.52% |
| 9 | history, century, modern, scientist, technology | 2.37% |
| 10 | environmental, change, land, climate, water | 7.63% |
| 11 | market, firms, labor, economic, financial | 2.94% |
| 12 | language, linguistic, speech, speaker, english | 5.50% |
| 13 | legal, law, right, violence, court | 2.53% |
| 14 | scholar, conference, international, scientist, issue | 4.14% |
| 15 | people, experiment, decision, test, game | 3.98% |
| 16 | behavior, social, conflict, individual, theory | 2.70% |
| 17 | archaeological, site, data, region, support | 4.38% |
| 18 | intellectual, impacts, merit, broader, support | 0.79% |
| 19 | technology, policy, organization, knowledge, innovation | 8.20% |
| 20 | human, species, genetic, evolution, data | 4.01% |
| 21 | policy, economic, country, international, prediction | 5.01% |
| 22 | training, graduate, undergraduate, grade, education | 2.16% |
| 23 | risk, income, health, effect, household | 2.27% |
| 24 | learning, cognitive, memory, human, knowledge | 2.83% |
| 25 | social, local, cultural, community, cross | 7.93% |
| 26 | spatial, objects, infant, action, categories | 0.66% |
| 27 | children, development, family, parent, adult | 2.10% |
| 28 | american, ethnic, migration, african, identity | 0.49% |
| 29 | health, medical, care, china, disease | 0.34% |
| 30 | altitude, pathology, health, adapt, risk | 0.32% |

Notes: asl means American sign language.

## Supplementary Note 5: Endogeneity of research funding

Endogeneity of research grants may lead to biased estimates. Supplementary Table 6 reports exogeneity tests for research funding across the main analysis categories. The tests indicate evidence of endogeneity in most, but not all, categories. The null hypothesis that funding is exogenous is rejected for the overall sample, the male and female subsamples, the male sample in female-dominated topics, and both gender groups in male-dominated topics. However, for female scholars in female-dominated topics, the null hypothesis is not rejected (p = 0.2939 for the robust score test and p = 0.2978 for the robust regression test). Therefore, for this subgroup, we report the non-IV regression estimates, while the remaining specifications are estimated using IV/2SLS.

**Supplementary Table 6. Results of exogeneity tests for research grants across different analysis categories.**

| Test | Overall | Male | Female | Female-dominated topics | | Male-dominated topics | |
|---|---|---|---|---|---|---|---|
| | | | | Male | Female | Male | Female |
| Robust score chi2(1) | 45.3723 | 25.0232 | 29.6073 | 6.6600 | 1.1015 | 23.2515 | 22.7800 |
| | (p=0.0000) | (p=0.0000) | (p=0.0000) | (p=0.0.0099) | (p=0.2939) | (p=0.0000) | (p=0.0000) |
| Robust regression *F* | 46.3329 | 25.2336 | 31.7736 | 8.8849 | 1.08434 | 23.4456 | 24.2956 |
| | (p=0.0000) | (p=0.0000) | (p=0.0000) | (p=0.0.0029) | (p=0.2978) | (p=0.0000) | (p=0.0000) |

Notes: H0: variables are exogenous.

# Supplementary Note 6: First-stage regression results of the 2SLS model

Displaying the results of the first-stage regression in 2SLS analysis is crucial for the IVs' study [53]. It allows us to assess the effectiveness of instrumental variables and the degree of endogeneity. The first-stage regressions assess whether instrumental variables are relevant predictors of research funding. A significant first-stage relationship supports instrument relevance but does not, by itself, establish the exclusion restriction. Therefore, we interpret the IV estimates under the maintained assumption that the instruments affect later academic performance only through their effect on funding, conditional on the included covariates and pre-funding academic performance [28]. In total we have three optional instrumental variables, isomorphism, dual appointment and SBE training. Since we grouped scholars into multiple categories and then ran IVs and 2SLS regressions, it is possible that the instrumental variables differed in their performance in controlling for endogeneity. Thus, at a minimum, we ensure that at least two instrument pairs in each regression are able to control for endogeneity in research funding. Having more instruments than the endogenous variables also ensures that we are measuring whether there is over-identification or under-identification of instruments [54].

The Supplementary Table 7 shows the 2SLS first-stage regression results for both males and females. We can find a significant effect of instrumental variables on research funding. Supplementary Table 8 and Supplementary Table 9 show the results of their 2SLS first-stage regressions under female-dominated research topics and male-dominated research topics, respectively. The instruments also have a significant effect on whether or not scholars receive research grants.

**Supplementary Table 7. First-stage regression results for male and female scholars.**

| **Male**<br>**Funding** | **Coefficient** | **Std. err.** | **t** | **P>t** | **[95% conf. interval]** | |
|---|---|---|---|---|---|---|
| Initial performance | -0.1988 | 0.0012 | -172.4000 | 0.0000 | -0.2011 | -0.1965 |
| Funding quota | -0.0161 | 0.0014 | -11.2200 | 0.0000 | -0.0189 | -0.0133 |
| Career age | -0.0025 | 0.0002 | -12.2300 | 0.0000 | -0.0029 | -0.0021 |
| Year | -0.0119 | 0.0005 | -24.1100 | 0.0000 | -0.0129 | -0.0109 |
| Fields' articles | -0.0217 | 0.0033 | -6.5500 | 0.0000 | -0.0282 | -0.0152 |
| Affiliations' articles | -0.1092 | 0.0048 | -22.5800 | 0.0000 | -0.1186 | -0.0997 |
| Fields' citations | 0.0216 | 0.0035 | 6.2000 | 0.0000 | 0.0147 | 0.0284 |
| Affiliations' citations | 0.1072 | 0.0043 | 24.9100 | 0.0000 | 0.0988 | 0.1156 |
| Employer reputation | -0.0005 | 0.0001 | -6.0600 | 0.0000 | -0.0007 | -0.0004 |
| USNEWS ranking | 0.0000 | 0.0000 | -0.0700 | 0.9450 | 0.0000 | 0.0000 |
| QS ranking | 0.0000 | 0.0000 | -7.3200 | 0.0000 | -0.0001 | 0.0000 |
| Isomorphism | 0.0020 | 0.0001 | 32.8500 | 0.0000 | 0.0019 | 0.0021 |
| SBE training | -0.0074 | 0.0012 | -6.1800 | 0.0000 | -0.0098 | -0.0051 |
| _cons | 24.4741 | 0.9916 | 24.6800 | 0.0000 | 22.5305 | 26.4176 |
| R-squared | 0.4864 | | | | | |
| Adj R-squared | 0.4862 | | | | | |
| Root MSE | 0.3533 | | | | | |

| **Female**<br>**Funding** | **Coefficient** | **Std. err.** | **t** | **P>t** | **[95% conf. interval]** | |
|---|---|---|---|---|---|---|
| Initial performance | -0.1858 | 0.0018 | -104.9600 | 0.0000 | -0.1893 | -0.1824 |
| Funding quota | -0.0139 | 0.0022 | -6.3200 | 0.0000 | -0.0183 | -0.0096 |
| Career age | -0.0037 | 0.0005 | -7.8800 | 0.0000 | -0.0046 | -0.0028 |
| Year | -0.0160 | 0.0008 | -21.2000 | 0.0000 | -0.0174 | -0.0145 |
| Fields' articles | -0.0250 | 0.0048 | -5.2500 | 0.0000 | -0.0343 | -0.0156 |
| Affiliations' articles | -0.1025 | 0.0088 | -11.6900 | 0.0000 | -0.1197 | -0.0853 |
| Fields' citations | 0.0248 | 0.0050 | 4.9600 | 0.0000 | 0.0150 | 0.0346 |
| Affiliations' citations | 0.0925 | 0.0077 | 12.0800 | 0.0000 | 0.0775 | 0.1075 |
| Employer reputation | -0.0001 | 0.0001 | -0.6400 | 0.5250 | -0.0003 | 0.0002 |
| USNEWS ranking | 0.0000 | 0.0000 | -0.0400 | 0.9670 | 0.0000 | 0.0000 |
| QS ranking | 0.0000 | 0.0000 | -4.6100 | 0.0000 | -0.0001 | 0.0000 |
| Isomorphism | 0.0021 | 0.0001 | 20.5500 | 0.0000 | 0.0019 | 0.0023 |
| Dual appointment | 0.1125 | 0.0461 | 2.4400 | 0.0150 | 0.0222 | 0.2029 |
| _cons | 32.6821 | 1.5151 | 21.5700 | 0.0000 | 29.7123 | 35.6519 |
| R-squared | 0.4249 | | | | | |
| Adj R-squared | 0.4244 | | | | | |
| Root MSE | 0.3680 | | | | | |

**Supplementary Table 8. First-stage regression results for male and female scholars in female-dominated research topics.**

| **Male**<br>**Funding** | **Coefficient** | **Std. err.** | **t** | **P>t** | **[95% conf. interval]** | |
|---|---|---|---|---|---|---|
| Initial performance | -0.1980 | 0.0042 | -46.7800 | 0.0000 | -0.2063 | -0.1897 |
| Funding quota | -0.0007 | 0.0051 | -0.1300 | 0.8960 | -0.0106 | 0.0093 |
| Career age | -0.0032 | 0.0006 | -5.5200 | 0.0000 | -0.0043 | -0.0021 |
| Year | -0.0052 | 0.0018 | -2.8900 | 0.0040 | -0.0087 | -0.0017 |
| Fields’ articles | -0.0267 | 0.0114 | -2.3500 | 0.0190 | -0.0490 | -0.0044 |
| Affiliations’ articles | -0.0633 | 0.0263 | -2.4100 | 0.0160 | -0.1148 | -0.0118 |
| Fields’ citations | 0.0277 | 0.0119 | 2.3200 | 0.0200 | 0.0043 | 0.0511 |
| Affiliations’ citations | 0.0750 | 0.0241 | 3.1200 | 0.0020 | 0.0278 | 0.1222 |
| Employer reputation | -0.0009 | 0.0003 | -3.0800 | 0.0020 | -0.0015 | -0.0003 |
| USNEWS ranking | 0.0000 | 0.0000 | -1.7900 | 0.0730 | -0.0001 | 0.0000 |
| QS ranking | -0.0001 | 0.0000 | -4.0100 | 0.0000 | -0.0001 | 0.0000 |
| Isomorphism | 0.0081 | 0.0010 | 7.8700 | 0.0000 | 0.0061 | 0.0101 |
| Dual appointment | 0.1520 | 0.0678 | 2.2400 | 0.0250 | 0.0189 | 0.2850 |
| SBE training | -0.0135 | 0.0046 | -2.9400 | 0.0030 | -0.0225 | -0.0045 |
| _cons | 10.8385 | 3.6158 | 3.0000 | 0.0030 | 3.7481 | 17.9289 |
| R-squared | 0.5220 | | | | | |
| Adj R-squared | 0.5192 | | | | | |
| Root MSE | 0.3437 | | | | | |

| **Female**<br>**Funding** | **Coefficient** | **Std. err.** | **t** | **P>t** | **[95% conf. interval]** | |
|---|---|---|---|---|---|---|
| Initial performance | -0.1824 | 0.0043 | -42.8700 | 0.0000 | -0.1907 | -0.1740 |
| Funding quota | 0.0152 | 0.0054 | 2.8300 | 0.0050 | 0.0047 | 0.0257 |
| Career age | -0.0064 | 0.0009 | -7.0000 | 0.0000 | -0.0082 | -0.0046 |
| Year | -0.0116 | 0.0018 | -6.4700 | 0.0000 | -0.0151 | -0.0081 |
| Fields’ articles | -0.0451 | 0.0117 | -3.8600 | 0.0000 | -0.0680 | -0.0222 |
| Affiliations’ articles | -0.0186 | 0.0226 | -0.8200 | 0.4110 | -0.0629 | 0.0257 |
| Fields’ citations | 0.0495 | 0.0122 | 4.0700 | 0.0000 | 0.0257 | 0.0733 |
| Affiliations’ citations | 0.0251 | 0.0197 | 1.2800 | 0.2020 | -0.0135 | 0.0637 |
| Employer reputation | -0.0006 | 0.0003 | -2.0100 | 0.0450 | -0.0013 | 0.0000 |
| USNEWS ranking | -0.0001 | 0.0000 | -2.7500 | 0.0060 | -0.0001 | 0.0000 |
| QS ranking | -0.0001 | 0.0000 | -3.1500 | 0.0020 | -0.0001 | 0.0000 |
| Isomorphism | 0.0047 | 0.0006 | 7.7600 | 0.0000 | 0.0035 | 0.0059 |
| Dual appointment | 0.4036 | 0.0655 | 6.1600 | 0.0000 | 0.2751 | 0.5320 |
| _cons | 23.5999 | 3.6071 | 6.5400 | 0.0000 | 16.5265 | 30.6733 |
| R-squared | 0.4628 | | | | | |
| Adj R-squared | 0.4598 | | | | | |
| Root MSE | 0.3625 | | | | | |

**Supplementary Table 9. First-stage regression results for male and female scholars in male-dominated research topics.**

| **Male**<br>**Funding** | **Coefficient** | **Std. err.** | **t** | **P>t** | **[95% conf. interval]** | |
|---|---|---|---|---|---|---|
| Initial performance | -0.1984 | 0.0012 | -165.0700 | 0.0000 | -0.2008 | -0.1961 |
| Funding quota | -0.0171 | 0.0015 | -11.1900 | 0.0000 | -0.0201 | -0.0141 |
| Career age | -0.0025 | 0.0002 | -11.5400 | 0.0000 | -0.0030 | -0.0021 |
| Year | -0.0122 | 0.0005 | -23.7300 | 0.0000 | -0.0132 | -0.0112 |
| Fields' articles | -0.0217 | 0.0035 | -6.2800 | 0.0000 | -0.0285 | -0.0149 |
| Affiliations' articles | -0.1101 | 0.0049 | -22.2500 | 0.0000 | -0.1198 | -0.1004 |
| Fields' citations | 0.0215 | 0.0036 | 5.9400 | 0.0000 | 0.0144 | 0.0286 |
| Affiliations' citations | 0.1073 | 0.0044 | 24.4100 | 0.0000 | 0.0987 | 0.1160 |
| Employer reputation | -0.0005 | 0.0001 | -5.3000 | 0.0000 | -0.0007 | -0.0003 |
| USNEWS ranking | 0.0000 | 0.0000 | 0.2700 | 0.7900 | 0.0000 | 0.0000 |
| QS ranking | 0.0000 | 0.0000 | -6.2600 | 0.0000 | -0.0001 | 0.0000 |
| Isomorphism | 0.0020 | 0.0001 | 32.5000 | 0.0000 | 0.0018 | 0.0021 |
| SBE training | -0.0074 | 0.0013 | -5.9100 | 0.0000 | -0.0099 | -0.0050 |
| _cons | 25.0971 | 1.0325 | 24.3100 | 0.0000 | 23.0733 | 27.1210 |
| R-squared | 0.4845 | | | | | |
| Adj R-squared | 0.4842 | | | | | |
| Root MSE | 0.3538 | | | | | |

| **Female**<br>**Funding** | **Coefficient** | **std. err.** | **t** | **P>t** | **[95% conf. interval]** | |
|---|---|---|---|---|---|---|
| Initial performance | -0.1872 | 0.0020 | -94.9800 | 0.0000 | -0.1910 | -0.1833 |
| Funding quota | -0.0147 | 0.0026 | -5.7600 | 0.0000 | -0.0197 | -0.0097 |
| Career age | -0.0030 | 0.0005 | -6.1900 | 0.0000 | -0.0039 | -0.0020 |
| Year | -0.0166 | 0.0008 | -19.5600 | 0.0000 | -0.0183 | -0.0149 |
| Fields' articles | -0.0227 | 0.0053 | -4.3000 | 0.0000 | -0.0330 | -0.0123 |
| Affiliations' articles | -0.1218 | 0.0101 | -12.0500 | 0.0000 | -0.1416 | -0.1020 |
| Fields' citations | 0.0217 | 0.0055 | 3.9200 | 0.0000 | 0.0108 | 0.0326 |
| Affiliations' citations | 0.1087 | 0.0088 | 12.4000 | 0.0000 | 0.0915 | 0.1259 |
| Employer reputation | -0.0001 | 0.0001 | -0.6600 | 0.5080 | -0.0004 | 0.0002 |
| USNEWS ranking | 0.0000 | 0.0000 | 1.0700 | 0.2840 | 0.0000 | 0.0000 |
| QS ranking | 0.0000 | 0.0000 | -3.9400 | 0.0000 | -0.0001 | 0.0000 |
| Isomorphism | 0.0021 | 0.0001 | 19.9100 | 0.0000 | 0.0019 | 0.0023 |
| SBE training | -0.0045 | 0.0020 | -2.2400 | 0.0250 | -0.0084 | -0.0006 |
| _cons | 33.9556 | 1.7071 | 19.8900 | 0.0000 | 30.6093 | 37.3018 |
| R-squared | 0.4254 | | | | | |
| Adj R-squared | 0.4247 | | | | | |
| Root MSE | 0.3666 | | | | | |

## Supplementary Note 7: Introduction to instrumental variables

Using the number of awards received by colleagues as an instrument to control for the endogeneity of funding has yielded promising initial findings [29]. We have further refined this instrument by incorporating consideration of colleagues' research topics. Our approach involved gathering data on the number of grants awarded to scholars' institutions across various research topics during different historical periods before applying for funding. Subsequently, this data was utilized to gauge the cumulative impact of institutional support available for specific topics. This process entails the dissemination of information and preferential attachment of resources [55,56]. Academic institutions gradually amass funding and reputation in particular research fields, thereby establishing authority within those domains [57]. This cumulative effect enables subsequent applicants to recognize the benefits of research funding for their own topics of inquiry. Additionally, subsequent applicants are better positioned to comprehend crucial information regarding research funding applications for these topics, such as assessment criteria and funding proposal tendencies. Moreover, they may glean insights from past research funding applications, thereby enhancing their own potential success rates.

Hence, the accumulation of funding at individual scholar and research topic levels not only reflects the research activity of institutions in specific fields but also furnishes subsequent researchers with a more conducive research environment and enhanced funding prospects. According to the theory of isomorphic imitation, we refer to this instrument as inter-scholar isomorphism [58,59]. We calculated the standardized values of historical awards for scholars on the same research topics at the same institution prior to applying for funding,

$$isomorphism\{u,q,i,l\}(h) = \left(f\{u,q,i,l\}(1), f\{u,q,i,l\}(2), \ldots, f\{u,q,i,l\}(h-l)\right) \quad (6)$$

where $isomorphism\{u,q,i,l\}(h)$ represents a vector consisting of grant counts for each year of research topic $q$ at scholar $i$'s institution $u$, covering the years before the year, $l$, in which the scholar applies for funding. $h$ represents the length of the observation window for our entire sample, which is 20 years. Considering the differences in the magnitude of grant counts across universities, we standardized $isomorphism\{u,q,i,l\}(h)$ by:

$$isomorphism_{norm} = \frac{isomorphism - \min(isomorphism)}{\max(isomorphism) - \min(isomorphism)} \tag{7}$$

The second instrument is the NSF-sponsored training activities related to social sciences conducted at universities across the United States. We compiled data on the number of NSF days conducted by the SBE division specifically targeting the proposal writing and promotion activities for social science research grants from 2000 to 2019. NSF days are organized sporadically at selected universities or colleges with the aim of enhancing the skills and knowledge required for submitting research proposals, thereby influencing the chances of successfully securing funding. Some studies suggest that such training significantly impacts the proposal-writing skills of grant applicants, which differ substantially from the writing skills required for publishing articles. This design is intended to capture variation in proposal-writing support that affects the probability of obtaining funding. We acknowledge, however, that training activities could also be correlated with institutional resources or research environments, so the exclusion restriction remains an identifying assumption rather than a directly testable fact. Hence, we introduced a binary variable to signify whether the scholar's institution had conducted NSF days within the three years preceding their grant submission. A value of 1 denotes that the university organized such training sessions before the scholar's grant application, while a value of 0 indicates otherwise.

The third instrumental variable is the dual appointment of American social science scholars in non-profit organizations. Dual appointments, also considered by some scholars as academic hegemony, are utilized by scholars as an instrument to control for the endogeneity of European Union and Italy research funding, respectively [28,60,61]. We introduce this instrument into research funding in the American social sciences. We retrieved and compiled data on scholars' dual appointments before applying for funding from the NSF and other non-profit academic organizations. In our dataset, from 1990 to 2019, 58 PIs served or had served as members of the NSF Board of Directors or as rotators in SBE departments, or as external experts for grant review meetings or programs; 62 PIs were elected as members of the Social Sciences and Humanities Division/American Philosophical Society (APS) in seven fields related to SBE funding, such as sociology, demography, economics, or linguistics; and 142 PIs were selected as members or administrative staff of the American Association for the Advancement of Science (AAAS) in social and economic sciences and their interdisciplinary fields. We set this instrument variable to 1 if the scholar held dual appointments in the aforementioned organizations before obtaining funding, and 0 otherwise.

The theoretical justification, specific design, and validity of the three instrumental variables can be found in our previous work [27].

## Supplementary Note 8: Alternative validity tests for instrumental variables

We provide summary statistics values for the 2SLS model's first-stage regressions, which help us to understand whether the instruments are weak instrumental variables or not. Supplementary Table 10 reports the first-stage diagnostic statistics. The first-stage F-statistics range from 22.6521 to 553.1060, with all p-values below 0.001. These results indicate that the instruments are strongly correlated with research funding in the first stage. Supplementary Table 11 further reports Shea's partial R-squared and the minimum eigenvalue statistics for each 2SLS specification. Generally, when the F-statistic exceeds 10, it indicates that the instrumental variables are not weak. Therefore, the three instrumental variables we selected are not weak. Additionally, in Supplementary Table 11, we provide the minimum eigenvalue statistic for all 2SLS models. Although Shea's partial R-sq. is less than 0.04, none of the regression's minimum eigenvalue statistics differ significantly from the F-statistic values and are all greater than 10.

**Supplementary Table 10. 2SLS model summary statistics for first-stage regressions.**

| Test | | Adjusted R-sq. | Partial R-sq. | Robust R-sq. | F | Prob>F |
|---|---|---|---|---|---|---|
| Overall | | 0.4660 | 0.4658 | 0.0306 | 505.9940 | 0.0000 |
| Male | | 0.4864 | 0.4862 | 0.0348 | 553.1060 | 0.0000 |
| Female | | 0.4249 | 0.4244 | 0.0227 | 216.9930 | 0.0000 |
| Female-dominated topics | Male | 0.5220 | 0.5192 | 0.0267 | 22.6521 | 0.0000 |
| | Female | 0.4628 | 0.4598 | 0.0314 | 48.5207 | 0.0000 |
| Male-dominated topics | Male | 0.4845 | 0.4842 | 0.0368 | 541.0120 | 0.0000 |
| | Female | 0.4254 | 0.4247 | 0.0265 | 203.7960 | 0.0000 |

**Supplementary Table 11. 2SLS model Shea's statistics for first-stage regressions.**

| Test | Shea's partial R-sq. | Shea's Adj. partial R-sq. | Minimum eigenvalue statistic | Endogenous regressors | Excluded instruments |
|---|---|---|---|---|---|
| Overall | 0.0306 | 0.0303 | 463.5080 | 1 | 3 |
| Male | 0.0348 | 0.0345 | 546.6470 | 1 | 2 |

| Female | | 0.0227 | 0.0219 | 164.6750 | 1 | 2 |
|---|---|---|---|---|---|---|
| Female-dominated topics | Male | 0.0267 | 0.0213 | 21.3433 | 1 | 3 |
| | Female | 0.0314 | 0.0266 | 38.6984 | 1 | 2 |
| Male-dominated topics | Male | 0.0368 | 0.0364 | 534.7560 | 1 | 2 |
| | Female | 0.0265 | 0.0255 | 153.8550 | 1 | 2 |

The estimated values of the 2SLS model using instrumental variables are often biased, leading to size distortion, which increases with weak instrumental variables. We further compare the minimum eigenvalue statistics in Supplementary Table 11 with the Stock-Yogo weak-instrument critical values reported in Supplementary Table 12. The smallest minimum eigenvalue statistic is 21.3433 for the male sample in female-dominated topics. This value exceeds the Stock-Yogo 10% maximal size critical value for specifications with three excluded instruments (19.93), and the remaining specifications also exceed the relevant critical values for their number of excluded instruments. These diagnostics do not indicate a weak-instrument problem in the reported 2SLS specifications.

**Supplementary Table 12. Weak instrumental variables test results.**

| No. of IVs | Test | 10% | 15% | 20% | 25% |
|---|---|---|---|---|---|
| 2 | 2SLS Size of nominal | 22.3 | 12.83 | 9.54 | 7.8 |
| | LIML Size of nominal | 6.4600 | 4.3600 | 3.6900 | 3.3200 |
| 3 | 2SLS Size of nominal | 19.93 | 11.59 | 8.75 | 7.25 |
| | LIML Size of nominal | 8.6800 | 5.3300 | 4.4200 | 3.9200 |

H0: Instruments are weak

**Supplementary Table 13. Over-identification tests for instrumental variables.**

| Test | Overall | Male | Female | Female-dominated topics | | Male-dominated topics | |
|---|---|---|---|---|---|---|---|
| | | | | Male | Female | Male | Female |
| Score chi2(2) | 0.4612 (p = 0.7940) | 0.4388 (p = 0.5077) | 0.4815 (p = 0.4877) | 0.7517 (p = 0.6867) | 0.1032 (p = 0.7480) | 0.4261 (p = 0.5139) | 0.0449 (p = 0.8322) |

Supplementary Table 13 reports over-identification tests for specifications with more excluded instruments than endogenous regressors. The non-significant p-values indicate that the over-identifying restrictions are not rejected. However, these tests do not prove that all instruments are exogenous, nor do they test under-identification; they only provide diagnostic evidence consistent with the maintained IV assumptions.

## Supplementary References

1. Nickell, D. S. *Guidebook for the Scientific Traveler: Visiting Physics and Chemistry Sites Across America*. (Rutgers University Press, 2010).

2. Freeman, L. A. *Longing for the Bomb: Oak Ridge and Atomic Nostalgia*. (University of North Carolina Press, 2015).

3. Sanderson, K. More women than ever are starting careers in science. *Nature* **596**, 177 (2021).

4. Casad, B. J. *et al.* U.S. women faculty in the social sciences also face gender inequalities. *Front. Psychol.* **13**, (2022).

5. Winslow, S. Gender inequality and time allocations among academic faculty. *Gend. Soc.* **24**, 769–793 (2010).

6. Simpkins, S., Estrella, G., Gaskin, E. & Kloberdanz, E. Latino parents' science beliefs and support of high school students' motivational beliefs: Do the relations vary across gender and familism values? *Soc. Psychol. Educ.* **21**, 1203–1224 (2018).

7. Trusz, S. Why do females choose to study humanities or social sciences, while males prefer technology or science? Some intrapersonal and interpersonal predictors. *Soc. Psychol. Educ.* **23**, 615–639 (2020).

8. Antón, S. C., Malhi, R. S. & Fuentes, A. Race and diversity in U.S. biological anthropology: A decade of AAPA initiatives. *Am. J. Phys. Anthropol.* **165**, 158–180 (2018).

9. Ceci, S. J. & Williams, W. M. Understanding current causes of women's underrepresentation in science. *Proc. Natl. Acad. Sci.* **108**, 3157–3162 (2011).

10. Grant, J., Burden, S. & Breen, G. No evidence of sexism in peer review. *Nature* **390**, 438 (1997).

11. Hutchins, H. M. & Kovach, J. V. ADVANCING women academic faculty in STEM careers: The role of critical HRD in supporting diversity and inclusion. *Adv. Dev. Hum. Resour.* **21**, 72–91 (2019).

12. Boyle *et al.* Gender balance: Women are funded more fairly in social science. *Nature* **525**, 181 (2015).

13. van der Lee, R. & Ellemers, N. Gender contributes to personal research funding success in The Netherlands. *Proc. Natl. Acad. Sci.* **112**, 12349–12353 (2015).

14. National Science Foundation. *National Center for Science and Engineering Statistics [NCSES]. Table 59. Median Annual Salaries of U.S. Residing Full-Time Employed Doctoral Scientists and Engineers in 4-Year Educational Institutions, by Field of Doctorate, Sex, and Faculty Rank: 2019* (2021).

15. Eagly, A. H. Do the social roles that women and men occupy in science allow equal access to publication? *Proc. Natl. Acad. Sci.* **117**, 5553–5555 (2020).

16. Lundberg, S. & Stearns, J. Women in economics: Stalled progress. *J. Econ. Perspect.* **33**, 3–22 (2019).

17. Head, M. G., Fitchett, J. R., Cooke, M. K., Wurie, F. B. & Atun, R. Differences in research funding for women scientists: a systematic comparison of UK investments in global infectious disease research during 1997-2010. *BMJ Open* **3**, e003362 (2013).

18. Burns, K. E. A., Straus, S. E., Liu, K., Rizvi, L. & Guyatt, G. Gender differences in grant and personnel award funding rates at the Canadian Institutes of Health Research based on research content area: A retrospective analysis. *PLoS Med.* **16**, e1002935 (2019).

19. Auriol, E., Friebel, G., Weinberger, A. & Wilhelm, S. Underrepresentation of women in the economics profession more pronounced in the United States compared to heterogeneous Europe. *Proc. Natl. Acad. Sci.* **119**, e2118853119 (2022).

20. Larrègue, J. & Pavie, A. Prestige at play: University hierarchies and the reproduction of funding inequalities. *Can. Rev. Sociol.* **63**, (2025).

21. Schweiger, G. *et al.* The costs of competition in distributing scarce research funds. *Proc. Natl. Acad. Sci.* **121**, e2407644121 (2024).

22. Goulden, M., Mason, M. A. & Frasch, K. Keeping women in the science pipeline. *Ann. Am. Acad. Pol. Soc. Sci.* **638**, 141–162 (2011).

23. Feldon, D. F. *et al.* Postdocs' lab engagement predicts trajectories of PhD students' skill development. *Proc. Natl. Acad. Sci.* **116**, 20910–20916 (2019).

24. Imbens, G. W. & Athey, S. Identification and inference in nonlinear difference-in-difference models. *Econometrica* **74**, 431–497 (2006).

25. De Chaisemartin, C. & D'Haultfoeuille, X. Fuzzy differences-in-differences. *Rev. Econ. Stud.* **85**, 999–1028 (2018).

26. Howell, A. Socio-economic impacts of scaling back a massive payments for ecosystem services programme in China. *Nat. Hum. Behav.* **6**, 1218–1225 (2022).

27. Ding, Y. & Bu, Y. Political hegemony, imitation isomorphism, and project familiarity: Instrumental variables to understand funding impact on scholar performance. *Quant. Sci. Stud.* 1–34 (2025) doi:10.1162/qss_a_00359.

28. Lawson, C., Geuna, A. & Finardi, U. The funding-productivity-gender nexus in science, a multistage analysis. *Res. Policy* **50**, 104182 (2021).

29. Kelchtermans, S., Neicu, D. & Veugelers, R. Off the beaten path: What drives scientists' entry into new fields? *Ind. Corp. Chang.* **31**, 654–680 (2022).

30. Thakuria, A., Deka, D. & Chakraborty, I. Comparative analysis of impact, collaboration, and prestige of open access and subscription based LIS journals using SJR and Scopus indicators. *Int. Inf. Libr. Rev.* **56**, 384–403 (2024).

31. Haans, R. F. J., Pieters, C. & He, Z.-L. Thinking about U: Theorizing and testing U- and inverted U-shaped relationships in strategy research. *Strateg. Manag. J.* **37**, 1177–1195 (2016).

32. Siegler, R. S. Five generalizations about cognitive development. *Am. Psychol.* **38**, 263–277 (1983).

33. Bjorklund, D. F., Miller, P. H., Coyle, T. R. & Slawinski, J. L. Instructing children to use memory strategies: Evidence of utilization deficiencies in memory training studies. *Dev. Rev.* **17**, 411–441 (1997).

34. Pauls, F., Macha, T. & Petermann, F. U-shaped development: An old but unsolved problem. *Front. Psychol.* **4**, (2013).

35. Pell, A. N. Fixing the Leaky Pipeline: Women Scientists in Academia. *J. Anim. Sci.* **74**, 2843–2848 (1996).

36. Sheltzer, J. M. & Smith, J. C. Elite male faculty in the life sciences employ fewer women. *Proc. Natl. Acad. Sci.* **111**, 10107–10112 (2014).

37. Lee, Y. & Won, D. Trailblazing women in academia: Representation of women in senior faculty and the gender gap in junior faculty's salaries in higher educational institutions. *Soc. Sci. J.* **51**, 331–340 (2014).

38. Jackson, R. Women in science: Plugging the leaky pipeline, how can we retain more women in science? *Scientifica* (2013).

39. Moss-Racusin, C. A., Dovidio, J. F., Brescoll, V. L., Graham, M. J. & Handelsman, J. Science faculty's subtle gender biases favor male students. *Proc. Natl. Acad. Sci.* **109**, 16474–16479 (2012).

40. Carvan, T. Why aren't women in science applying for grants? *College of Science, Australian National University* (2022).

41. Bird, D. K. S. Do women publish fewer journal articles than men? Sex differences in publication productivity in the social sciences. *Br. J. Sociol. Educ.* **32**, 921–937 (2011).

42. Ross, M. B. *et al.* Women are credited less in science than men. *Nature* **608**, 135–145 (2022).

43. Malisch, J. L. *et al.* In the wake of COVID-19, academia needs new solutions to ensure gender equity. *Proc. Natl. Acad. Sci.* **117**, 15378–15381 (2020).

44. Buckee, C. *et al.* Women in science are battling both Covid-19 and the patriarchy. *Times higher education* (2020).

45. Currie, J., Thiele, B. & Harris, P. *Gendered Universities in Globalized Economies: Power, Careers, and Sacrifices*. (Lexington Books, 2002).

46. Beaudry, C. & Larivière, V. Which gender gap? Factors affecting researchers' scientific impact in science and medicine. *Res. Policy* **45**, 1790–1817 (2016).

47. Stoop, D., Cobo, A. & Silber, S. Fertility preservation for age-related fertility decline. *Lancet* **384**, 1311–1319 (2014).

48. Madsen, E. B. & Aagaard, K. Concentration of danish research funding on individual researchers and research topics: Patterns and potential drivers. *Quant. Sci. Stud.* **1**, 1159–1181 (2020).

49. King, G. & Nielsen, R. Why Propensity Scores Should Not Be Used for Matching. *Polit. Anal.* **27**, 435–454 (2019).

50. Dong, X. *et al.* Beyond correlation: Towards matching strategy for causal inference in Information Science. *J. Inf. Sci.* 1–14 (2021) doi:10.1177/0165551520979868.

51. Blei, D. M., Ng, A. Y., Jordan, M. I. & Lafferty, J. Latent dirichlet allocation. *J. Mach. Learn. Res.* **3**, 993–1022 (2003).

52. Hausman, J. Specification tests in econometrics. *Econometrica* **46**, 1251–1271 (1978).

53. Lal, A., Lockhart, M., Xu, Y. & Zu, Z. How Much Should We Trust Instrumental Variable Estimates in Political Science? Practical Advice based on Over 60 Replicated Studies. *Polit. Methods Quant. Methods eJournal* (2021).

54. Murray, M. P. Avoiding Invalid Instruments and Coping with Weak Instruments. *J. Econ. Perspect.* **20**, 111–132 (2006).

55. Bol, T., de Vaan, M. & van de Rijt, A. The Matthew effect in science funding. *Proc. Natl. Acad. Sci. U. S. A.* **115**, 4887–4890 (2018).

56. Antonelli, C. & Crespi, F. The 'Matthew effect' in R&D public subsidies: The Italian evidence. *Technol. Forecast. Soc. Change* **80**, 1523–1534 (2013).

57. Cyrenne, P. & Grant, H. University decision making and prestige: An empirical study. *Econ. Educ. Rev.* **28**, 237–248 (2009).

58. Wedlin, L. The Role of Rankings in Codifying a Business School Template: Classifications, Diffusion and Mediated Isomorphism in Organizational Fields. *Wiley-Blackwell Eur. Manag. Rev.* (2007).

59. Budd, R. Isomorphic tensions and anxiety in UK social science doctoral provision. *Policy Rev. High. Educ.* (2023).

60. Cattaneo, M., Horta, H. & Meoli, M. Dual appointments and research collaborations outside academia: evidence from the European academic population. *Stud. High. Educ.* 2066–2080 (2018) doi:10.1080/03075079.2018.1492534.

61. National Science Board. Chapter 7 science and technology: Public attitudes and understanding. in *Science and Engineering indicators 2018* (2018).